\documentclass[preprint2]{aastex61}

\usepackage{amsmath}
\usepackage[capitalise]{cleveref} 
\usepackage{textcomp} % \textmu

\submitjournal{ApJL}

\shorttitle{Wind--driven chemistry of HD~209458b}
\shortauthors{Drummond et al.}

\begin{document}

\title{Observable signatures of wind--driven chemistry with a fully
  consistent three dimensional radiative hydrodynamics model of HD~209458b}

\correspondingauthor{Benjamin Drummond}
\email{b.drummond@exeter.ac.uk}

\author{B. Drummond}
\affil{Astrophysics Group, University of Exeter, EX4 2QL, Exeter, UK}
\author{N. J. Mayne}
\affiliation{Astrophysics Group, University of Exeter, EX4 2QL, Exeter, UK}
\author{J. Manners}
\affiliation{Met Office, Exeter, EX1 3PB, UK}
\affiliation{Astrophysics Group, University of Exeter, EX4 2QL, Exeter, UK}
\author{A. L. Carter}
\affiliation{Astrophysics Group, University of Exeter, EX4 2QL, Exeter, UK}
\author{I. A. Boutle}
\affiliation{Met Office, Exeter, EX1 3PB, UK}
\affiliation{Astrophysics Group, University of Exeter, EX4 2QL, Exeter, UK}
\author{I. Baraffe}
\affiliation{Astrophysics Group, University of Exeter, EX4 2QL, Exeter, UK}
\affiliation{Univ Lyon, Ens de Lyon, Univ Lyon1, CNRS, CRAL, UMR5574, F-69007, Lyon, France}
\author{\'E. H\'ebrard}
\affil{Astrophysics Group, University of Exeter, EX4 2QL, Exeter, UK}
\author{P. Tremblin}
\affiliation{Maison de la simulation, CEA, CNRS, Univ. Paris-Sud, UVSQ, Université Paris-Saclay, 91191 Gif-Sur-Yvette, France}
\author{D. K. Sing}
\affiliation{Astrophysics Group, University of Exeter, EX4 2QL, Exeter, UK}
\author{D. S. Amundsen}
\affiliation{Department of Applied Physics and Applied Mathematics, Columbia University, New York, NY 10025, USA}
\affiliation{NASA Goddard Institute for Space Studies, New York, NY 10025, USA}
\author{D. Acreman}
\affiliation{Astrophysics Group, University of Exeter, EX4 2QL, Exeter, UK}

% ABSTRACT
\begin{abstract}
  We present a study of the effect of wind-driven advection on the
  chemical composition of hot Jupiter atmospheres using a
  fully-consistent 3D hydrodynamics, chemistry and radiative transfer
  code, the Met Office Unified Model (UM). Chemical modelling of
  exoplanet atmospheres has primarily been restricted to 1D models
  that cannot account for 3D dynamical processes. In this work we
  couple a chemical relaxation scheme to the UM to account for the
  chemical interconversion of methane and carbon monoxide. This is
  done consistently with the radiative transfer meaning that
  departures from chemical equilibrium are included in the heating
  rates (and emission) and hence complete the feedback between the
  dynamics, thermal structure and chemical composition. In this letter
  we simulate the well studied atmosphere of HD~209458b. We find that
  the combined effect of horizontal and vertical advection leads to an
  increase in the methane abundance by several orders of magnitude; directly opposite to
  the trend found in previous works. Our results demonstrate the need
  to include 3D effects when considering the chemistry of hot Jupiter
  atmospheres. We calculate transmission and emission spectra, as well
  as the emission phase curve, from our simulations. We conclude that
  gas-phase non-equilibrium chemistry is unlikely to explain the
  model--observation discrepancy in the 4.5\,{\textmu m} {\it
    Spitzer}/IRAC channel. However, we highlight other spectral
  regions, observable with the James Webb Space Telescope, where
  signatures of wind-driven chemistry are more prominant.
\end{abstract}

%%%%%%%%%%%%%%%%%%%%%%%%
% INTRODUCTION
%%%%%%%%%%%%%%%%%%%%%%%%
\section{Introduction}

Gas-phase chemical composition is crucial in determining the circulation, temperature and observable properties of planetary atmospheres. The opacity, determined by the composition, controls the heating which, in turn, drives the dynamics.

Chemical modelling of exoplanet atmospheres has largely been restricted to 1D chemical kinetics codes \citep[e.g.][]{Moses2011,Venot2012,Zahnle2014,DruTB16,TsaLG2017}, showing that vertical transport-induced quenching can lead to departures from chemical equilibrium.

Hot Jupiter atmospheres have large horizontal temperature gradients, due to intense irradiation of the dayside, that drive fast horizontal wind velocities. Therefore non-equilibrium chemistry due to horizontal transport has been speculated to be at least as important as vertical mixing in hot Jupiter atmospheres \citep{Showman2009,Moses2011} and has been suggested to explain the model--observation discrepancy found in thermal phase curves for several hot Jupiters \citep{KnuLF12,ZelLK14,WonKK16}.

Using a 3D atmosphere model with temperature and chemical relaxation schemes, \citet{CooS06} investigated the effect of wind-driven advection on the chemistry of methane, carbon monoxide and water. They found vertical quenching from a deep level of the atmosphere leads to a horizontally uniform composition at lower pressures. \citet{CooS06} conclude that {\it vertical} (not horizontal) mixing is dominant in determining the chemical composition.

\citet{AguPV14} considered a 1D chemical kinetics model with a time-varying temperature profile (a ``pseudo-2D model''), representing a column of atmosphere rotating around the equator. \citet{AguPV14} conclude that {\it horizontal} quenching is more important than vertical quenching, with the nightside atmosphere being ``contaminated'' by the composition of the dayside.

We investigate the effect of wind-driven advection on the atmospheric composition using a 3D general circulation model consistently coupled with a flexible radiative transfer scheme and a chemical relaxation scheme. We simulate the atmosphere of HD~209458b and find significant quantitative and qualitative differences with both \citet{CooS06} and \citet{AguPV14}. We conclude that a combination of horizontal and vertical transport determines the 3D chemical composition of hot Jupiter atmospheres. We also identify particular wavelength regions where wind-driven chemistry can cause observable differences in the transmission and emission spectra, compared with what is predicted assuming chemical equilibrium.

%%%%%%%%%%%%%%%%%%%%%%%%
% MODEL DESCRIPTION
%%%%%%%%%%%%%%%%%%%%%%%%
\section{Model description}

We use the Met Office Unified Model (UM), based on a dynamical core that solves the deep-atmosphere, non-hydrostatic Navier--Stokes equations \citep{WooSW14,MayBA14b,MayDB17}. The model uses the open-source SOCRATES\footnote{\url{https://code.metoffice.gov.uk/trac/socrates}} radiative transfer scheme \citep{Edw96,EdwS96} to calculate the radiative flux and heating rates in each column \citep{AmuBT14,AmuMB16,AmuTM17}. The UM has previously been applied to the atmospheres of hot Jupiters \citep{MayBA14,AmuMB16,MayDB17}, sub-Neptunes \citep{DruMB18} and terrestrial exoplanets \citep{BouMD17}.

% CHEMICAL RELAXATTION SCHEME%
\subsection{Chemical relaxation scheme}
\label{section:chem_relax}

Previous UM simulations of hydrogen--dominated atmospheres \citep[e.g.][]{AmuMB16} have assumed local chemical equilibrium. Here we relax this assumption by consistently coupling a chemical relaxation scheme to the UM, following the method of \citet{CooS06}.

The chemical relaxation scheme parameterises the kinetic interconversion of chemical species. In this study we focus on the major gas-phase absorbers methane, carbon monoxide and water. The material derivative of the carbon monoxide mole fraction $f_{\rm CO}$ is
\begin{equation}
\label{eq:co_relax}
	\frac{D f_{\rm CO}}{D t} = \frac{\partial f_{\rm CO}}{\partial t} + \frac{u}{r\cos\phi}\frac{\partial f_{\rm CO}}{\partial \lambda}  + \frac{v}{r}\frac{\partial f_{\rm CO}}{\partial \phi} + w\frac{\partial f_{\rm CO}}{\partial r}, 
\end{equation}
where $t$ is time, $r$ is radius, $\phi$ is latitude, $\lambda$ is longitude and $u$, $v$ and $w$ are the zonal, meridional and vertical wind velocities, respectively. The first term on the right describes changes of $f_{\rm CO}$ with time due to local processes (such as chemical conversion) while the remaining terms describe the advection processes in each direction.

Using the chemical relaxation method we write
\begin{equation}
\label{eq:relax}
  \frac{\partial f_{\rm CO}}{\partial t} = - \frac{f_{\rm CO}-f_{\rm CO,eq}}{\tau_{\rm chem}},
\end{equation}
where $f_{\rm CO,eq}$ is the chemical equilibrium mole fraction of carbon monoxide and $\tau_{\rm chem}$ is a chemical timescale. The advection of the tracer is performed by the extensively tested, semi-implicit, semi-Lagrangian advection scheme of the UM \citep[see][for details]{WooSW14}.

\cref{eq:relax} describes the ``relaxation'' of $f_{\rm CO}$ toward a prescribed equilibrium profile $f_{\rm CO,eq}$ on a given timescale $\tau_{\rm chem}$. We use the same calculation of $\tau_{\rm chem}$ as \citet{CooS06}.

Assuming all carbon is contained in carbon monoxide and methane and all oxygen is contained in carbon monoxide and water the mole fractions of methane $f_{\rm CH_4}$ and water $f_{\rm H_2O}$ can be calculated from $f_{\rm CO}$ via mass balance,
\begin{alignat}{2}
  f_{\rm CH_4} + f_{\rm CO} = A_{\rm C} = 4.57\times10^{-4}  \label{eq:mass_balance_c} \\
  f_{\rm H_2O} + f_{\rm CO} = A_{\rm O} = 8.31\times10^{-4}  \label{eq:mass_balance_o}
\end{alignat}
where $A_{\rm C}$ and $A_{\rm O}$ are constants based on the solar elemental abundances \citep{Asplund2009}\footnote{Several previous publications \citep{TreAM15,DruTB16,GoyMS18,DruMB18} erroneously referred to \citet{Caffau2011} for the elemental abundances. The correct reference for the elemental abundances is \citet{Asplund2009}.}.

The above assumption is a good one for the high temperature, solar composition atmospheres of hot Jupiters, where other carbon and oxygen containing species are generally several orders of magnitude less abundant \citep[e.g.][]{Burrows1999,Moses2011}. However, the assumption will break down for above solar metallicities where species such as HCN and CO$_2$ become more abundant \citep[e.g.][]{MosLV13}. We have compared the equilibrium methane mole fraction calculated via the mass balance approach with that from a full Gibbs energy minimisation method and found negligible differences.

We use the same method as \citet{CooS06} to calculate the equilibrium mole fractions of methane, carbon monoxide and water, assuming fixed mole fractions of hydrogen $f_{\rm H_2}=0.853$ and helium $f_{\rm He}=0.145$. 

We have tested our model by reproducing the results of \citet{CooS06}, using a similar temperature relaxation scheme. Qualitatively we find very similar results: wind-driven advection leads to a vertically and horizontally uniform composition, with almost all carbon in carbon monoxide, for pressures less than 1 bar. In chemical equilibrium we also reproduce the large methane abundance on the nightside. Quantitative differences are small ($\sim1\%$) but expected to exist due to many differences between the two models: e.g. model discretisation, resolution, numerical methods and diffusion schemes.

\subsection{Model setup and simulation parameters}

We use the same simulation parameters for HD~209458b as \citet{AmuMB16} with 144 points in longitude and 90 in latitude, and 66 vertical levels and dynamical and radiative timesteps of 30\,s and 150\,s, respectively. We refer the reader to \citet{AmuBT14,AmuMB16,AmuTM17} for a complete description of the radiative transfer method, calculation of the opacities and sources of the line-lists. 

Simulated observables are calculated directly from the UM, using a higher spectral resolution than when computing the heating rates \citep[see][]{BouMD17}. The transmission spectrum is computed from the 3D model using all the columns on the nightside of the limb. Columns are treated independently using spherical shell geometry and the resultant spectra are averaged to give a global mean. A full description is given in \citet{LinMM18}.

We use the advected $f_{\rm CO}$, and corresponding $f_{\rm CH_4}$ and $f_{\rm H_2O}$ from mass balance, to calculate the opacity due to these species. Therefore, the radiative transfer is computed consistently with the advected chemistry of these molecules, allowing feedback between the local composition, the thermal structure and the circulation.

For other chemical species included as opacity sources \citep{AmuMB16} we assume chemical equilibrium abundances. The equilibrium mole fraction of molecular nitrogen is calculated in an analogous way to carbon monoxide \citep{CooS06}, using the net reaction
\begin{equation}
  {\rm N_2} + {\rm 3H_2} \leftrightarrow {\rm 2NH_3}.
\end{equation}
The equilibrium mole fraction of ammonia is then found via mass balance with N$_2$,

\begin{equation}
  2f_{\rm N_2} + f_{\rm NH_3} = A_{\rm N} = 1.16\times10^{-4}.
\end{equation}

The mole fractions of the alkali species are estimated using the same parameterisations as \citet{AmuMB16}.

We present results from two simulations which are identical, except that the first assumes local chemical equilibrium while the second includes advection and chemical relaxation of $f_{\rm CO}$. We refer to these simulations thoughout the rest of this letter simply as the ``equilibrium'' and ``relaxation'' simulations.

The simulations are initialised with zero wind velocities and a horizontally uniform pressure-temperature ($P$--$T$) profile, as in \citet{AmuMB16}. The carbon monoxide tracer is initialised to chemical equilibrium corresponding to the initial $P$--$T$ profile, as in \citet{CooS06}. 

We integrate the model for 1000 Earth days, to compare with \citet{CooS06}, by which point the atmosphere has reached a steady state for pressures less than $10^6$ Pa. The deep atmosphere, on the other hand, continues to evolve \citep[e.g.][]{MayBA14,AmuMB16,MayDB17} after this time and requires integration times that are infeasible with current models.

%%%%%%%%%%%%%%%%%%%%%%%%
% DYNAMICS, THERMAL STRUCTURE AND COMPOSITION
%%%%%%%%%%%%%%%%%%%%%%%%
\section{Dynamics, thermal structure and composition}

% CH4 Mole fractions
\begin{figure*}

  \begin{center}
    \includegraphics[width=0.45\textwidth]{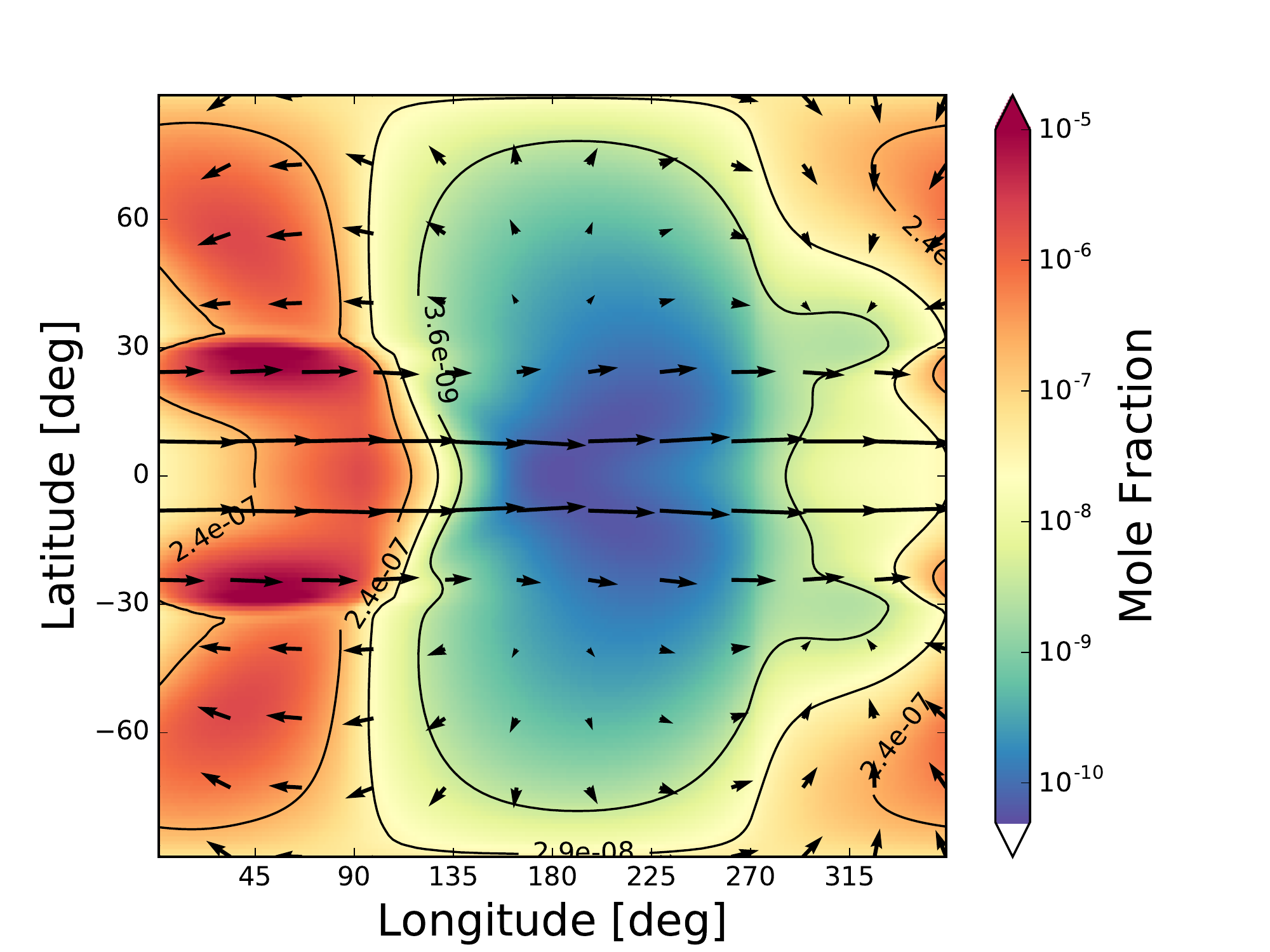}
    \includegraphics[width=0.45\textwidth]{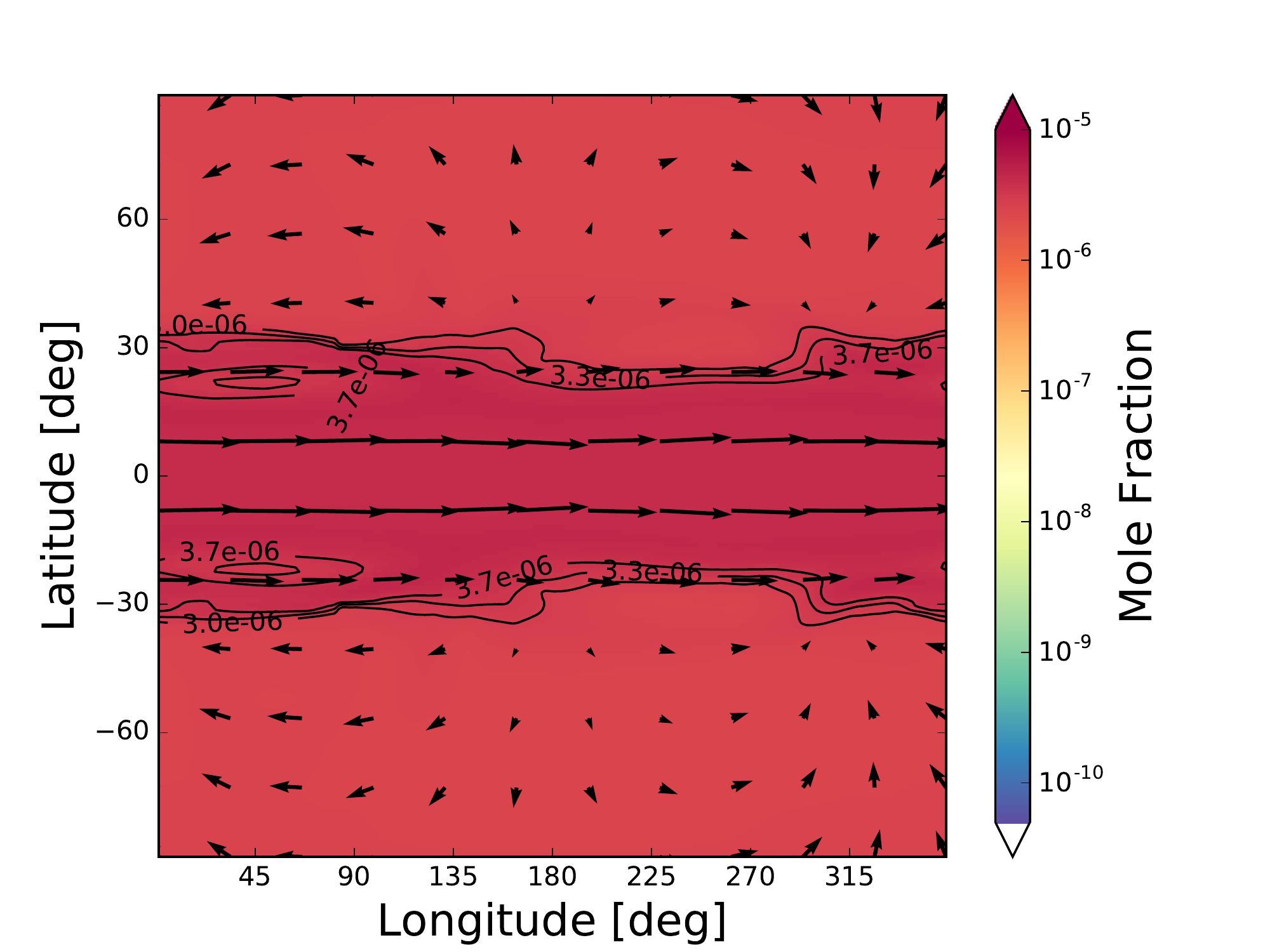} \\
    \includegraphics[width=0.45\textwidth]{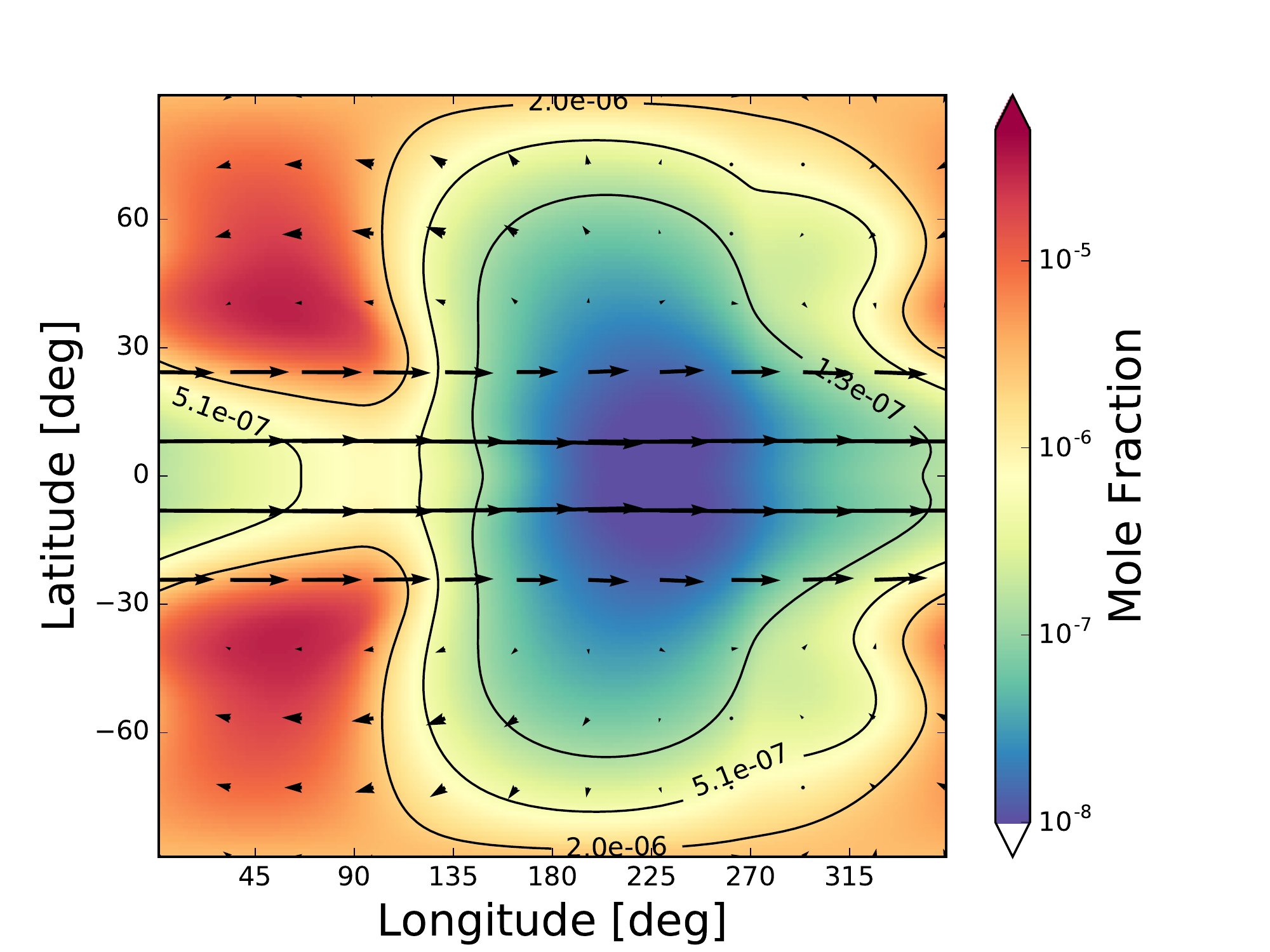}
    \includegraphics[width=0.45\textwidth]{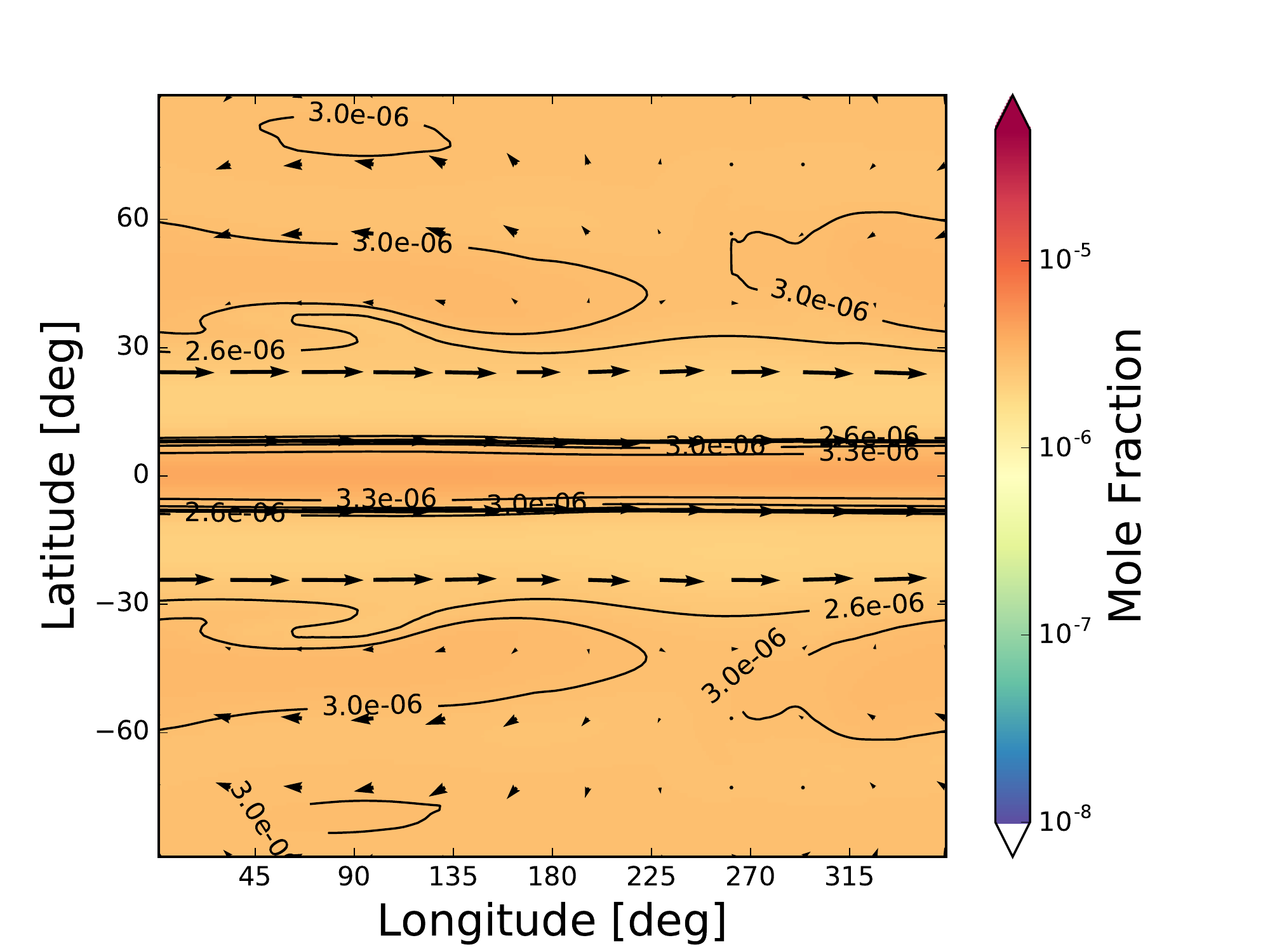} \\
    \includegraphics[width=0.45\textwidth]{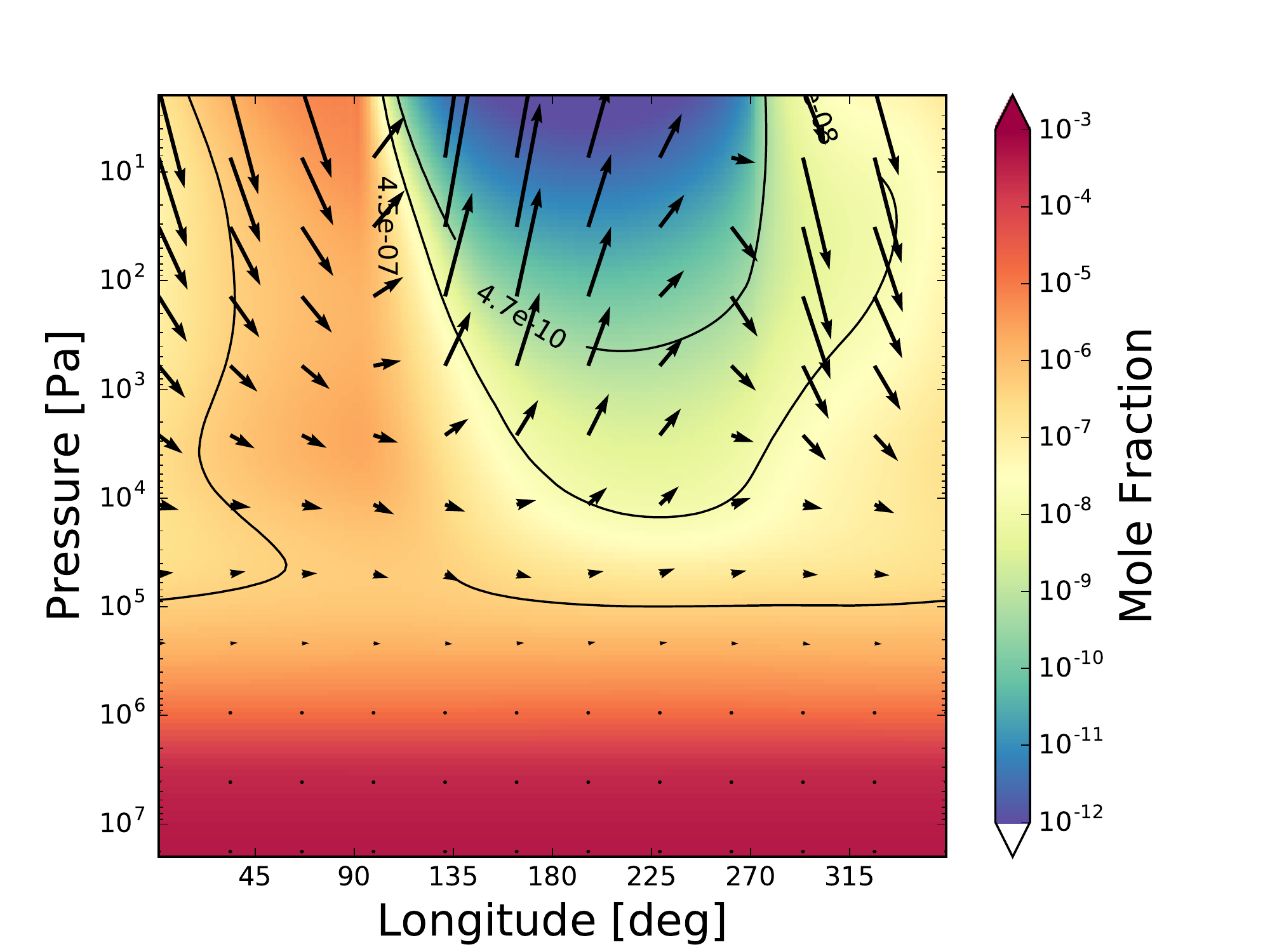}
    \includegraphics[width=0.45\textwidth]{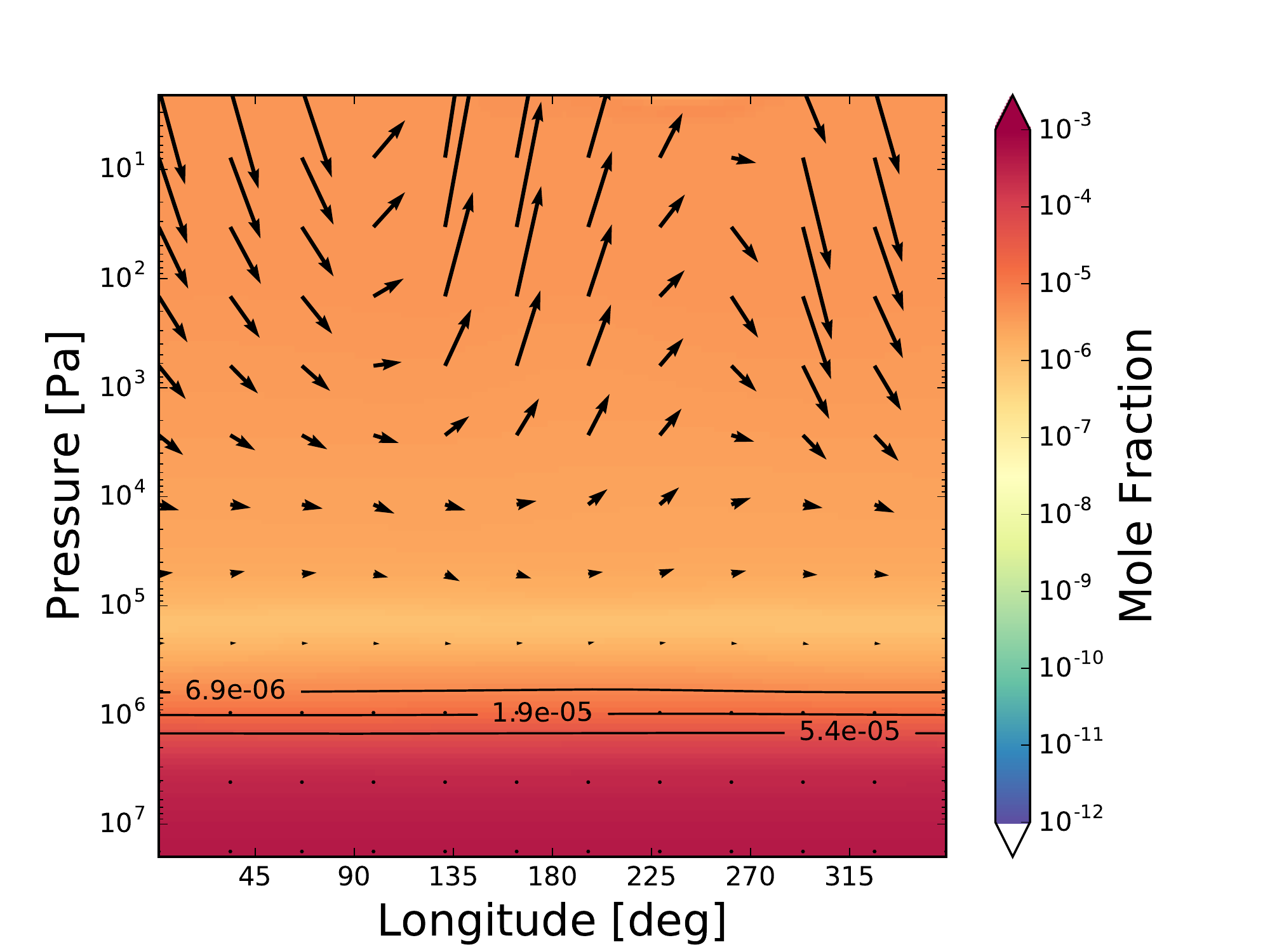}
  \end{center}

%\centering
%\gridline{\fig{fig1.pdf}{0.45\textwidth}{(a) $1\times10^2$ Pa}
%\fig{fig2.pdf}{0.45\textwidth}{(b) $1\times10^2$ Pa}}
%\gridline{\fig{fig3.pdf}{0.45\textwidth}{(c) $1\times10^4$ Pa}
%\fig{fig4.pdf}{0.45\textwidth}{(d) $1\times10^4$ Pa}}
%\gridline{\fig{fig5.pdf}{0.45\textwidth}{(e) $\bar{\phi}\left(\pm20^{\circ}\right)$}
%\fig{fig6.pdf}{0.45\textwidth}{(f) $\bar{\phi}\left(\pm20^{\circ}\right)$}}

\caption{{\it Left}: chemical equilibrium mole fractions of methane. {\it Right}: chemical relaxation mole fractions of methane. {\it Top} and {\it middle} are isobars of 1$\times10^2$ and 1$\times10^4$ Pa, respectively, with horizontal wind vectors and {\it bottom} an area-weighted meridional mean between $\pm20$ degrees latitude.}
\label{figure:ch4}
\end{figure*}

\begin{figure*}

  \begin{center}
       \includegraphics[width=0.45\textwidth]{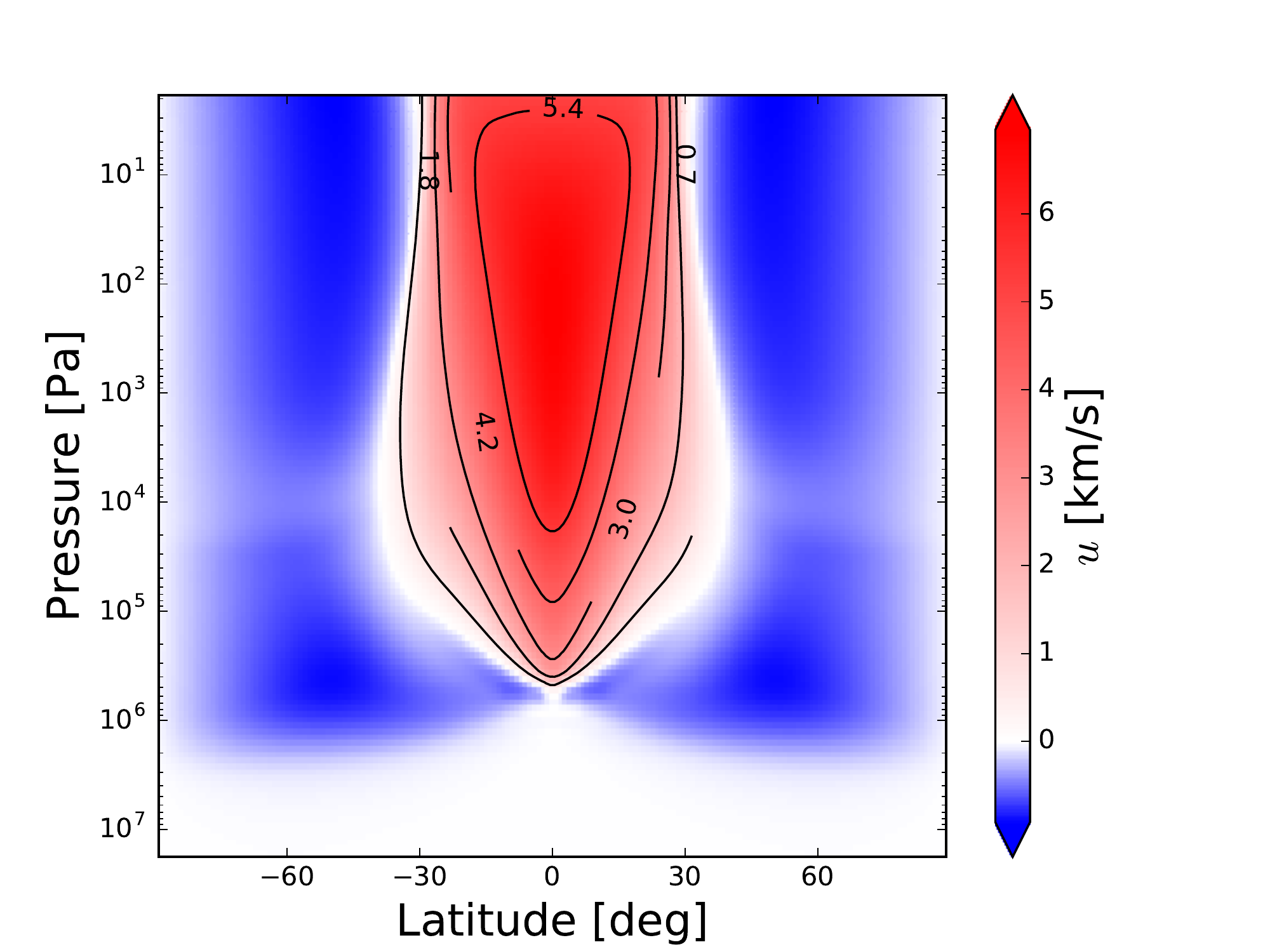}
    \includegraphics[width=0.45\textwidth]{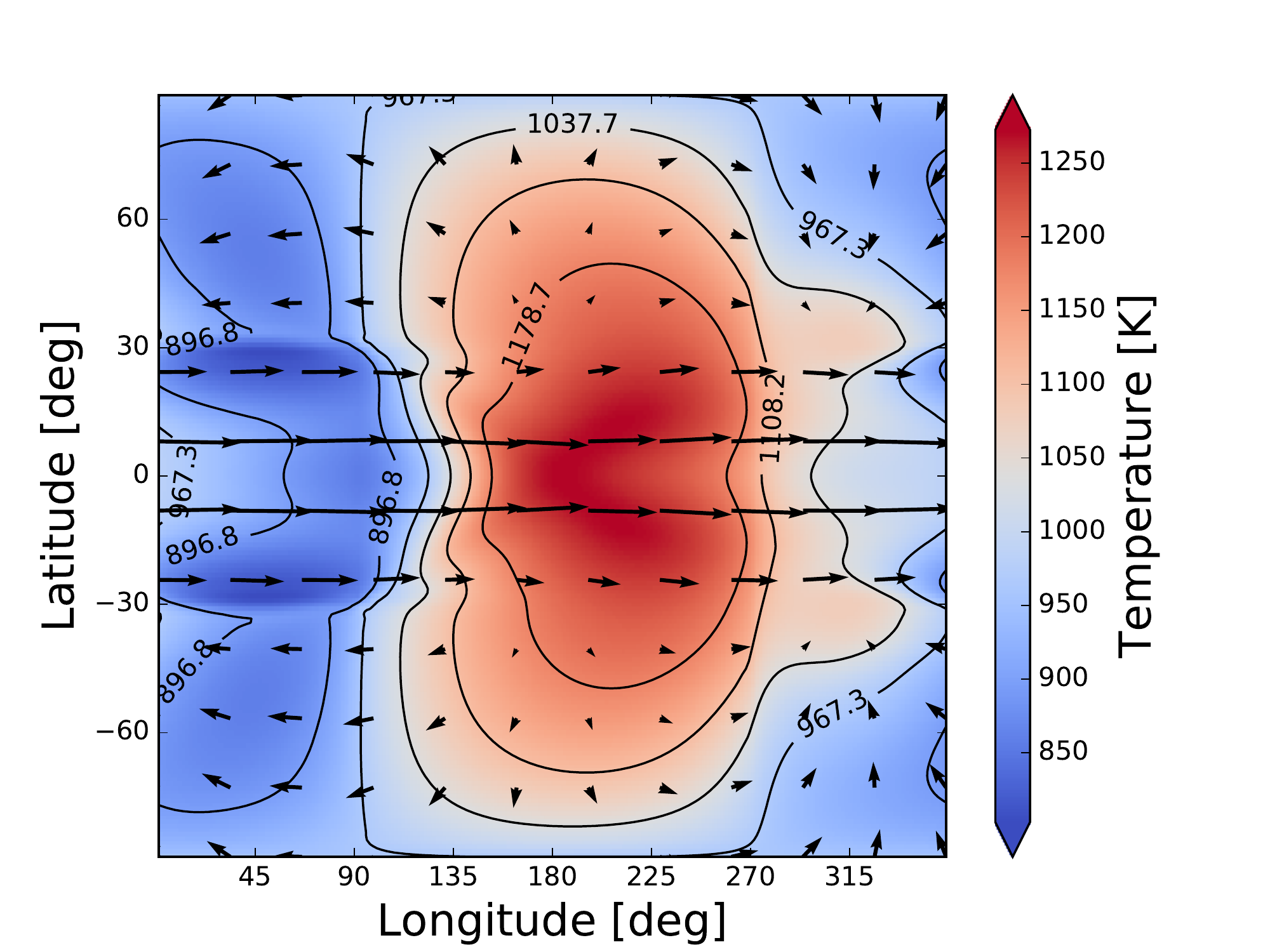} \\
    \includegraphics[width=0.45\textwidth]{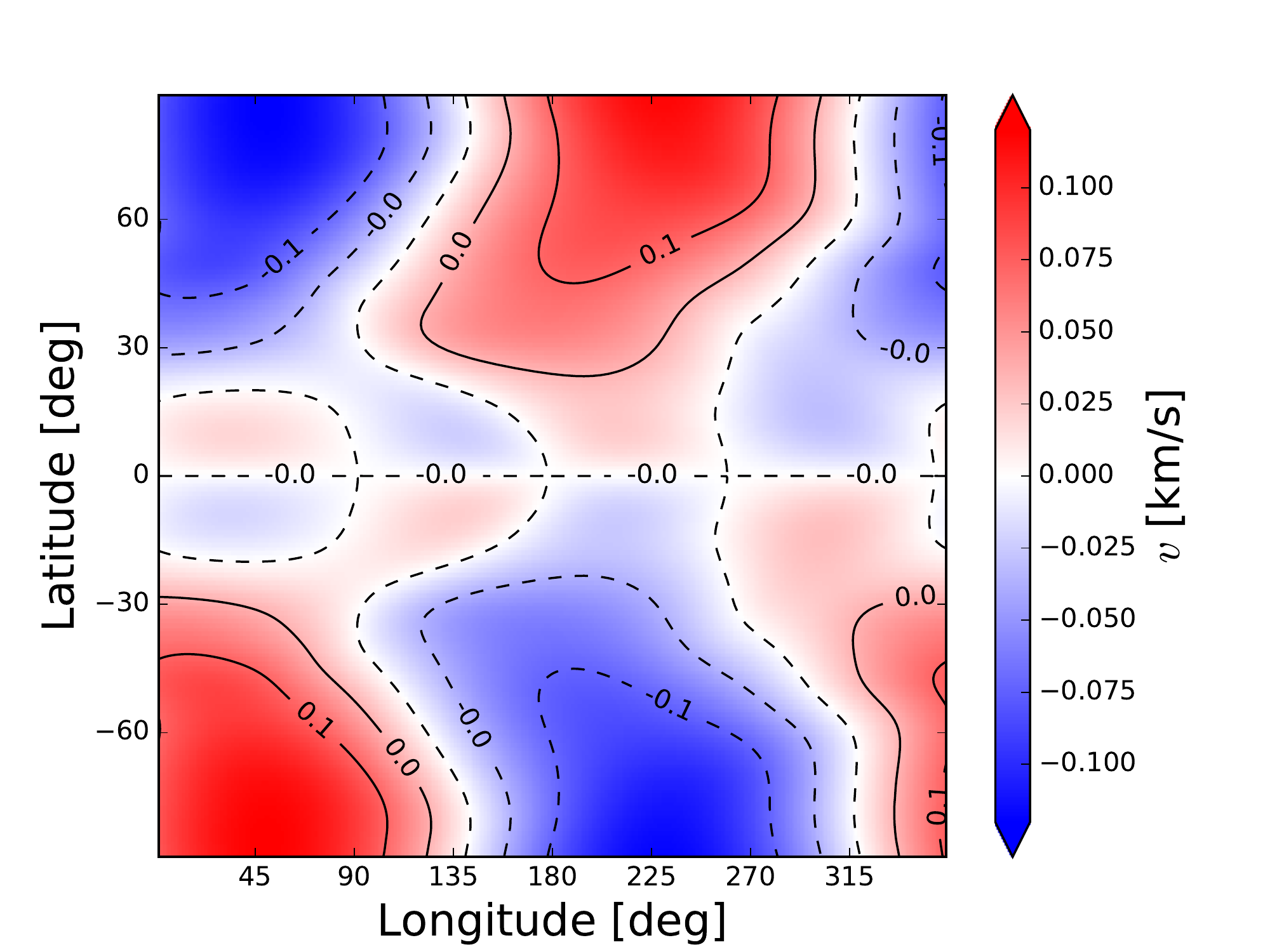}
    \includegraphics[width=0.45\textwidth]{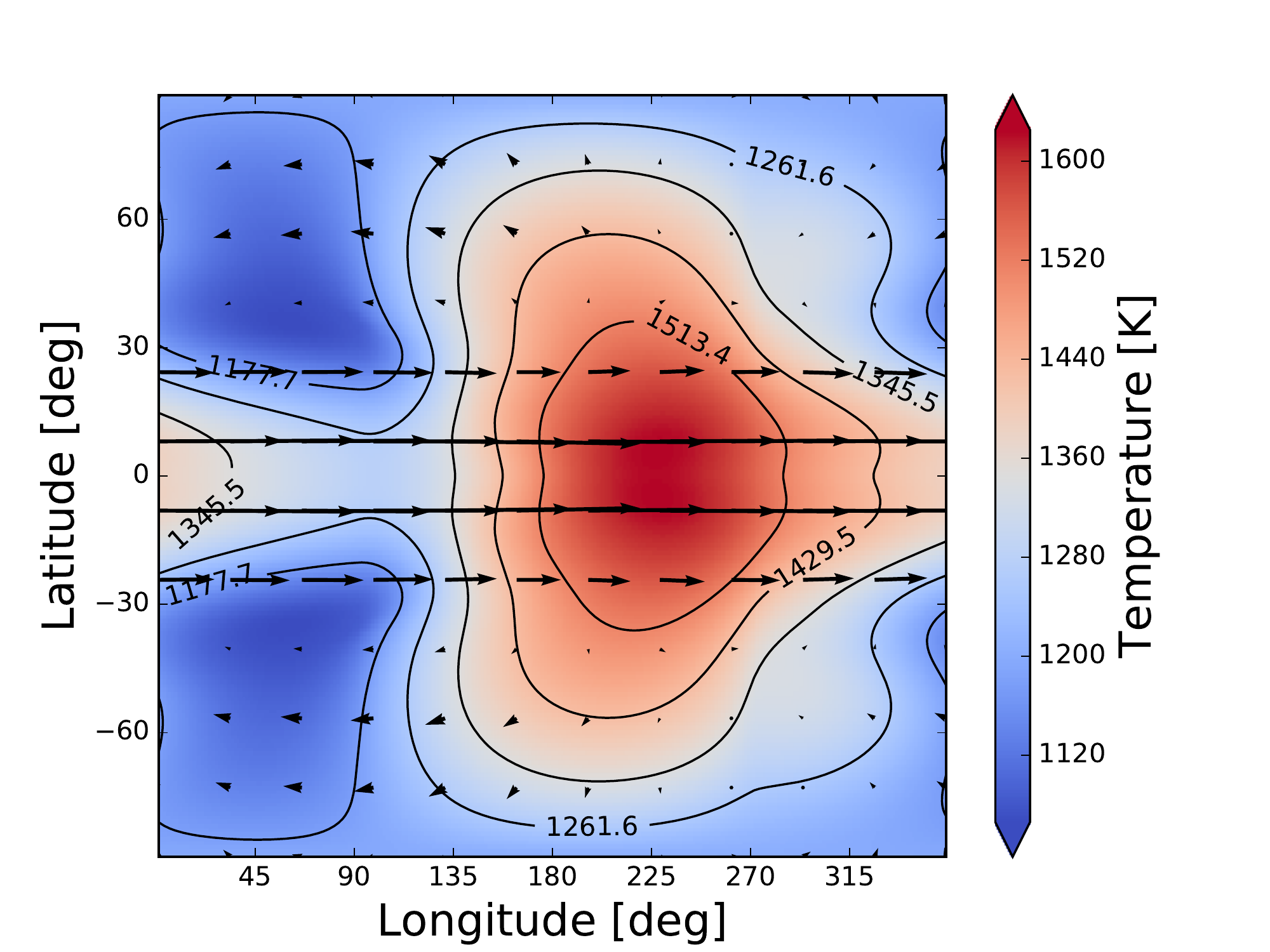} \\
    \includegraphics[width=0.45\textwidth]{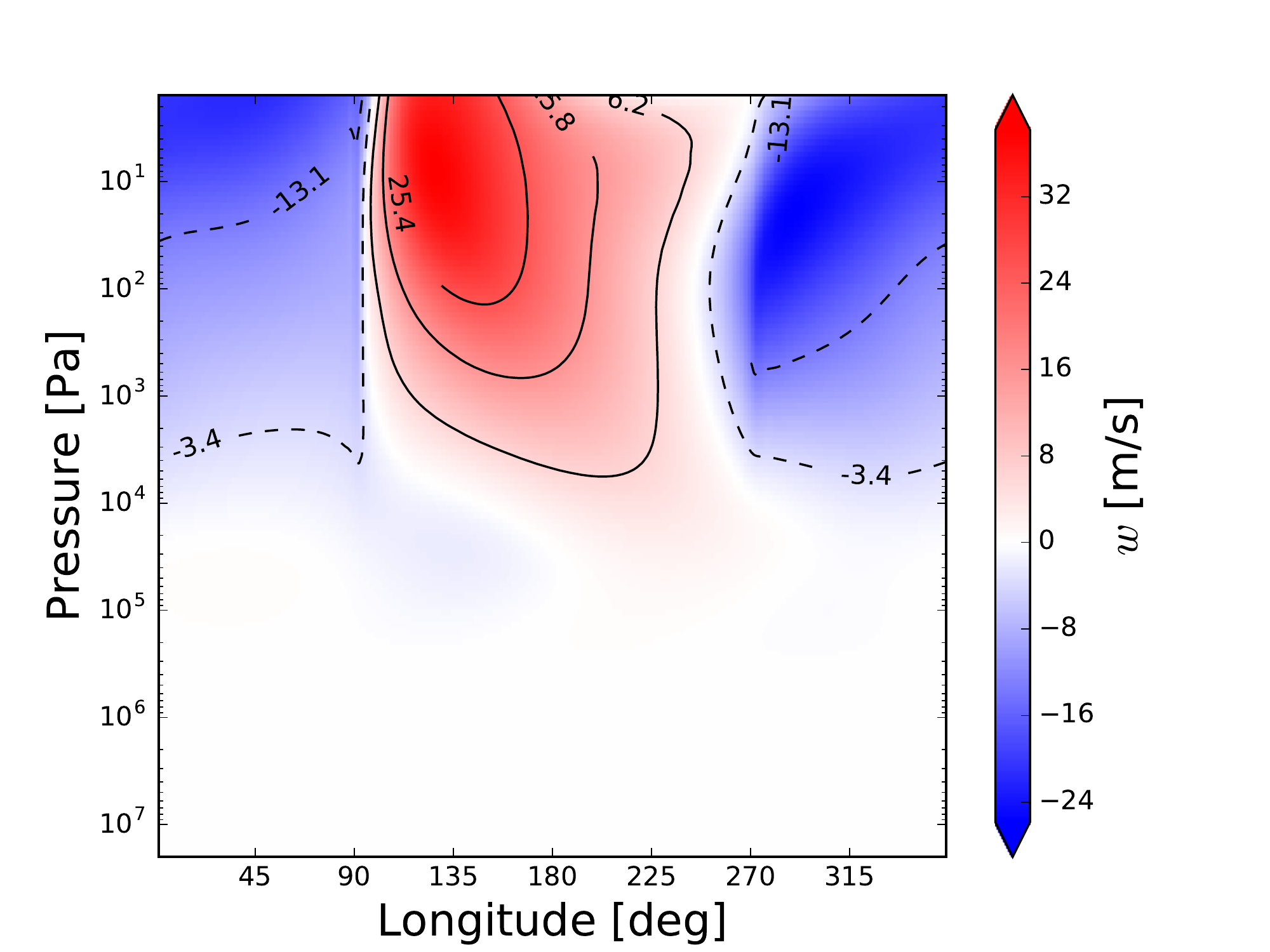}
    \includegraphics[width=0.45\textwidth]{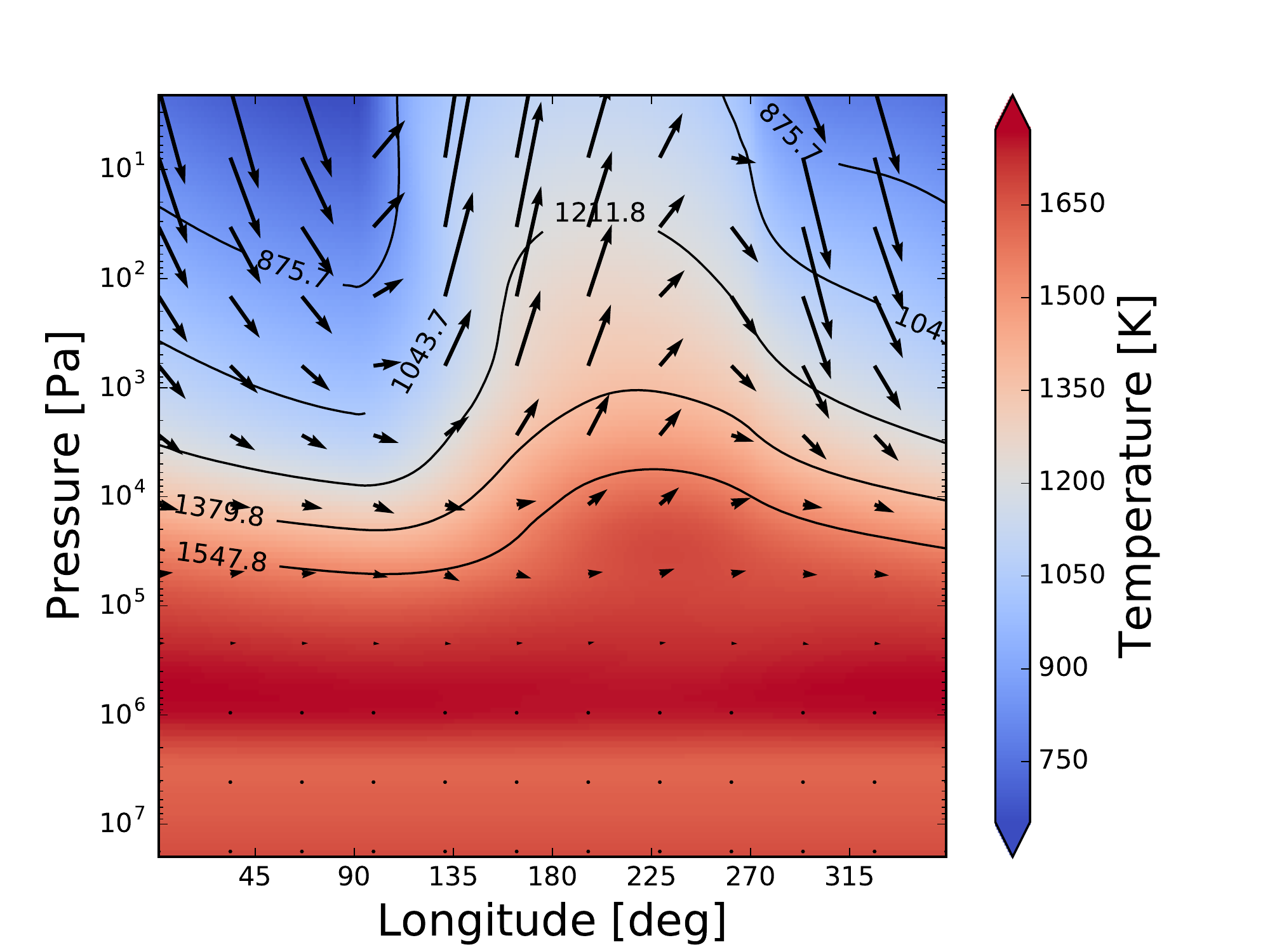}
  \end{center}

\caption{{\it Left}: wind velocities, showing (top) the zonal-mean of the zonal wind, (middle) the meridional wind at $5\times10^4$ Pa and (bottom) an  area--weighted meridional mean ($\pm20^{\circ}$) of the vertical wind. {\it Right}: temperatures, showing (top, middle) the $1\times10^2$ and $1\times10^4$ Pa isobars, respectively, with horizontal wind vectors and (bottom) an area-weighted meridional mean ($\pm20^{\circ}$) of the temperature with vertical-zonal wind vectors. Results are shown from the relaxation simulation.}
  \label{figure:winds_temp}
\end{figure*}

\cref{figure:ch4} compares the mole fraction of methane between the equilibrium and relaxation simulations. \cref{figure:winds_temp} shows the dynamical and thermal structure from the relaxation simulation, which are very similar to the results of \citet{AmuMB16}.

The equilibrium abundance of methane clearly traces the temperature structure (compare \cref{figure:ch4} \& \cref{figure:winds_temp}), as the chemistry is only dependent on the local temperature and pressure, for a given elemental composition. The equilibrium methane mole fraction varies with latitude and longitude by several orders of magnitude with larger abundances where the atmosphere is cooler.

% Temperature and Winds

Advection homogenises the methane abundance, both horizontally and vertically, across a large pressure range. Generally, the methane abundance in the relaxation simulation is larger than that in the equilibrium simulation. The winds that drive this homogenisation (\cref{figure:winds_temp}) are characterised by a strong equatorial zonal jet, upwelling on the dayside and downwelling on the nightside and alternating regions of poleward and equatorward flow in the meridional direction.

% EQUATORIAL MOLE FRACTIONS AND TIMESCALES
\begin{figure*}
  \begin{center}
    \includegraphics[width=0.45\textwidth]{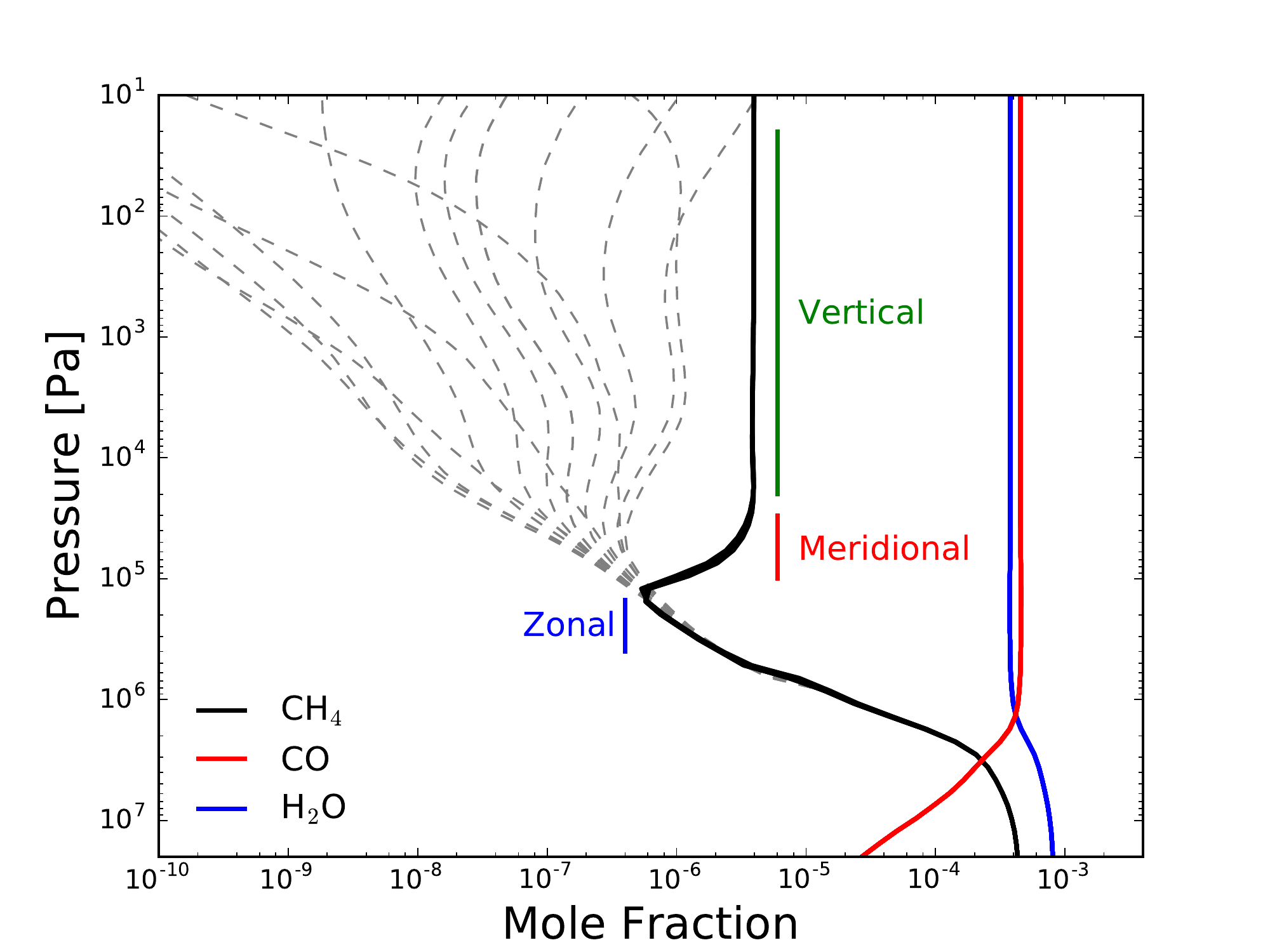}
    \includegraphics[width=0.45\textwidth]{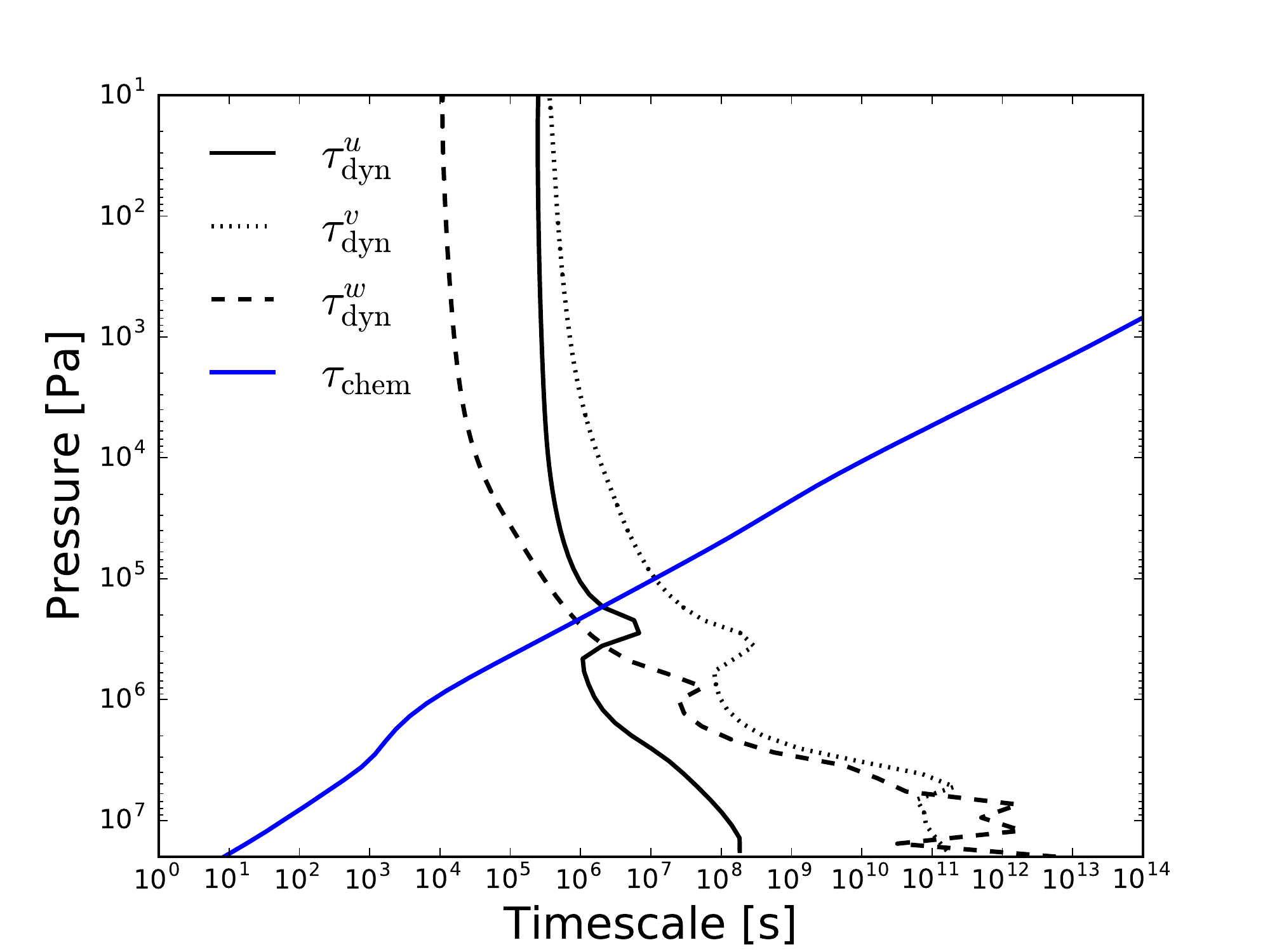}
  \end{center}

  \caption{{\it Left}: vertical abundance profiles of methane, carbon monoxide and water from the relaxation (solid) and equilibrium (dashed) simulations at a series of longitude points around the equator ($\phi=0^{\circ}$). {\it Right}: estimated dynamical and chemical timescales.}
  \label{figure:profiles}
\end{figure*}

\cref{figure:profiles} shows vertical profiles of the mole fractions of methane, carbon monoxide and water for a series of longitude points around the equator. In the relaxation simulation, methane is homogenised vertically and/or horizontally, depending on the pressure. The mole fractions of both carbon monoxide and water, which are approximately constant with both pressure (for $P>10^6$ Pa) and longitude in equilibrium, are not significantly effected.

Comparing the relaxation and equilibrium simulations, moving from high to low pressures, the abundance of methane deviates from equilibrium at $\sim3\times10^5$ Pa. Over a relatively narrow pressure range the methane abundance becomes uniform with longitude, with a value corresponding to the profile with the smallest equilibrium methane mole fraction. Here, {\it zonal} transport is driving the chemistry away from equilibrium.

Between $10^5$ and $10^4$ Pa the methane abundance increases with decreasing pressure, to values significantly above those shown in the equilibrium simulation. This is due to {\it meridional} transport of gas from higher latitudes, where methane is more abundant in equilibrium. The predominant meridional component of the wind velocity throughout the atmosphere is poleward, however small regions of equatorward flow, at low latitudes, act to transport gas toward the equator (see \cref{figure:winds_temp}).

For pressures less than $10^4$ Pa the methane abundance becomes approximately uniform with pressure due to {\it vertical} quenching, with a quenched mole fraction of $\sim$$\,3\times10^{-6}$; larger than the values predicted by chemical equilibrium.

In summary, at the equator, we find the mole fraction of methane is sequentially effected by zonal, meridional and vertical transport, moving from high to low pressures. To first order, we find that vertical quenching is the most important process, and determines the methane abundance for all pressures less than $P\sim10^4$ Pa. {\it However, importantly, the abundance at the vertical quench point is determined by horizontal transport.} This shows the importance of including 3D transport when considering transport-induced quenching of chemical species in exoplanet atmospheres.

Since we include interaction between the dynamics, radiative transfer and chemistry, we can quantify the effect of wind-driven chemistry on the circulation and temperature fields. We find modest differences ($\sim$\,$1\%$). This is not surprising, in this particular case, as the important absorbers water and carbon monoxide show negligible departure from chemical equilibrium. In addition, though methane varies from the equilibrium mole fraction by several orders of magnitude, its abundance remains small ($f_{\rm CH_4}\leq3\times10^{-6}$).

\cref{figure:profiles} shows the estimated dynamical and chemical timescales from our simulations. The dynamical timescale is split into the zonal $\tau_{\rm dyn}^u$, meridional $\tau_{\rm dyn}^v$ and vertical $\tau_{\rm dyn}^w$ components, which we estimate as 
\begin{alignat}{3}
	\tau_{\rm dyn}^u = \frac{L}{u} = \frac{2\pi R_{\rm p}}{u} \\
 	\tau_{\rm dyn}^v = \frac{L}{v}  = \frac{\pi R_{\rm p}}{2v} \\
	\tau_{\rm dyn}^w = \frac{H}{w},
\end{alignat}
where $L$ is the relevant horizontal length scale, $R_{\rm p}$ is the planetary radius and $H$ is the vertical scale height.  The relevant chemical timescale here is for the interconversion of carbon monoxide and methane, discussed in Section 2.1. We have assumed dayside averages of $u$, $w$ and $T$, for $\tau_{\rm dyn}^u$, $\tau_{\rm dyn}^w$ and $\tau_{\rm chem}$, respectively. For $\tau_{\rm dyn}^v$ we assumed the dayside average of $v$ over the northern hemisphere ($0^{\circ}<\phi<90^{\circ}$) only. 

Where $\tau_{\rm chem}<\tau_{\rm dyn}$ we expect chemical equilibrium to hold, while for  $\tau_{\rm chem}>\tau_{\rm dyn}$ we instead expect advection to drive the local composition. Each of the directional components of the dynamical timescale cross $\tau_{\rm chem}$ at approximately $10^5$ Pa, which agrees well with the pressure level at which methane diverges from equilibrium. We note that the timescales presented here serve as order of magnitude estimates, which depend on the assumed horizontal or vertical length scale and on the sampled wind velocity and temperature. In reality, it is the ratio of the {\it local} chemical and dynamical timescale that is of importance. The timescales shown in \cref{figure:profiles} are therefore not expected to predict the precise pressure level of the zonal, meridional and vertical quench points in our simulation.

\begin{figure*}
  \begin{center}
    \includegraphics[width=0.45\textwidth]{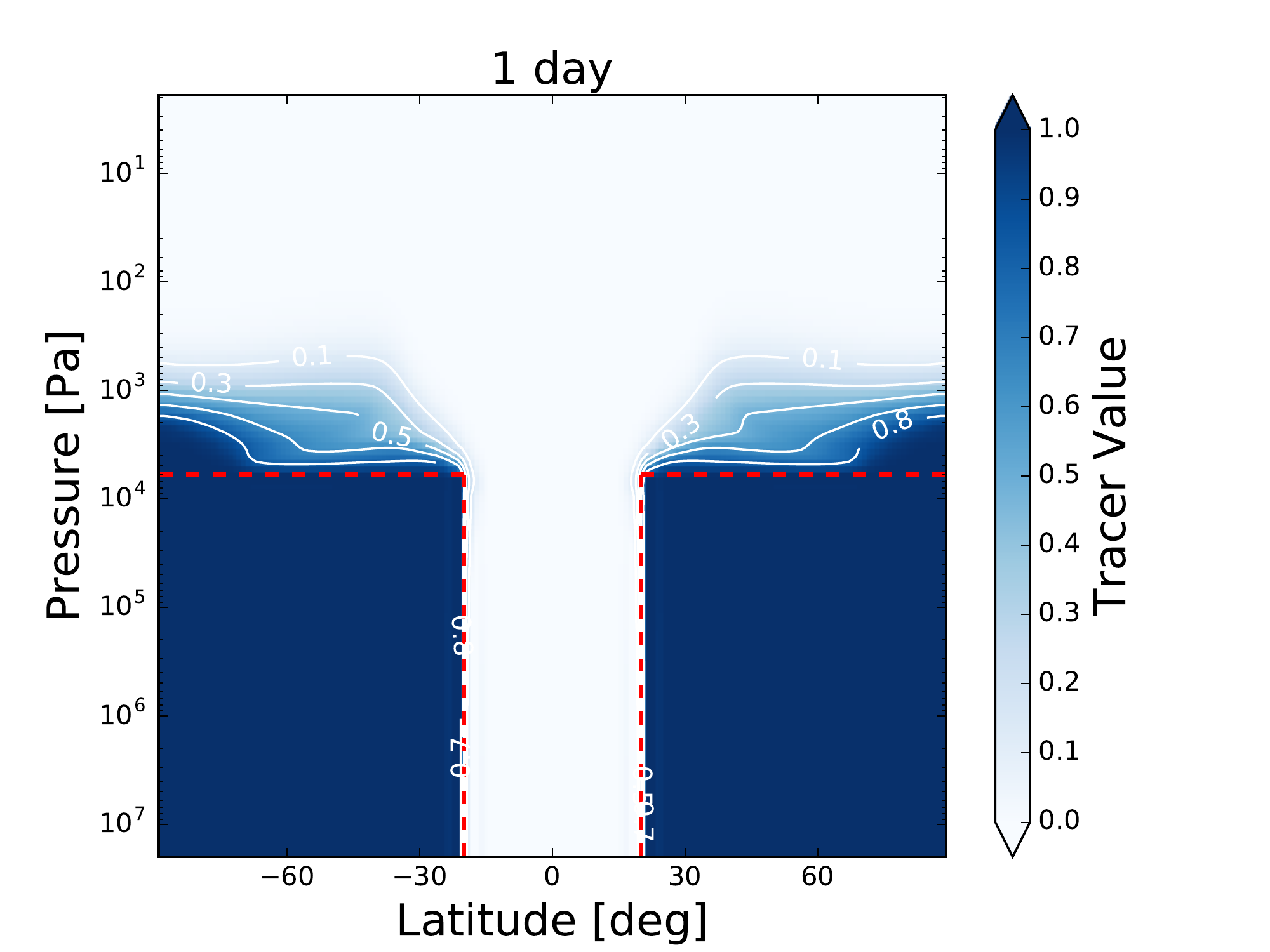}
    \includegraphics[width=0.45\textwidth]{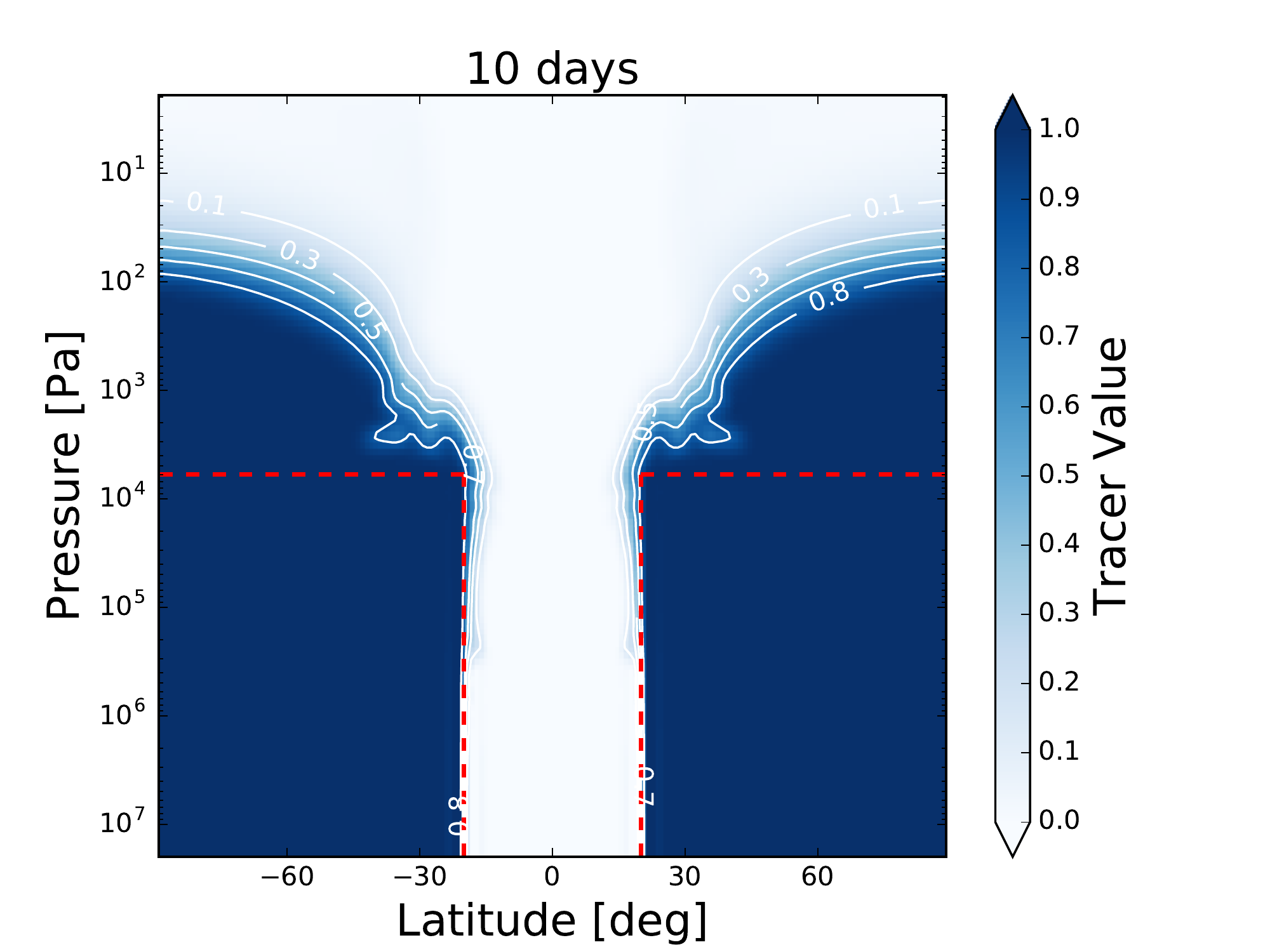} \\
    \includegraphics[width=0.45\textwidth]{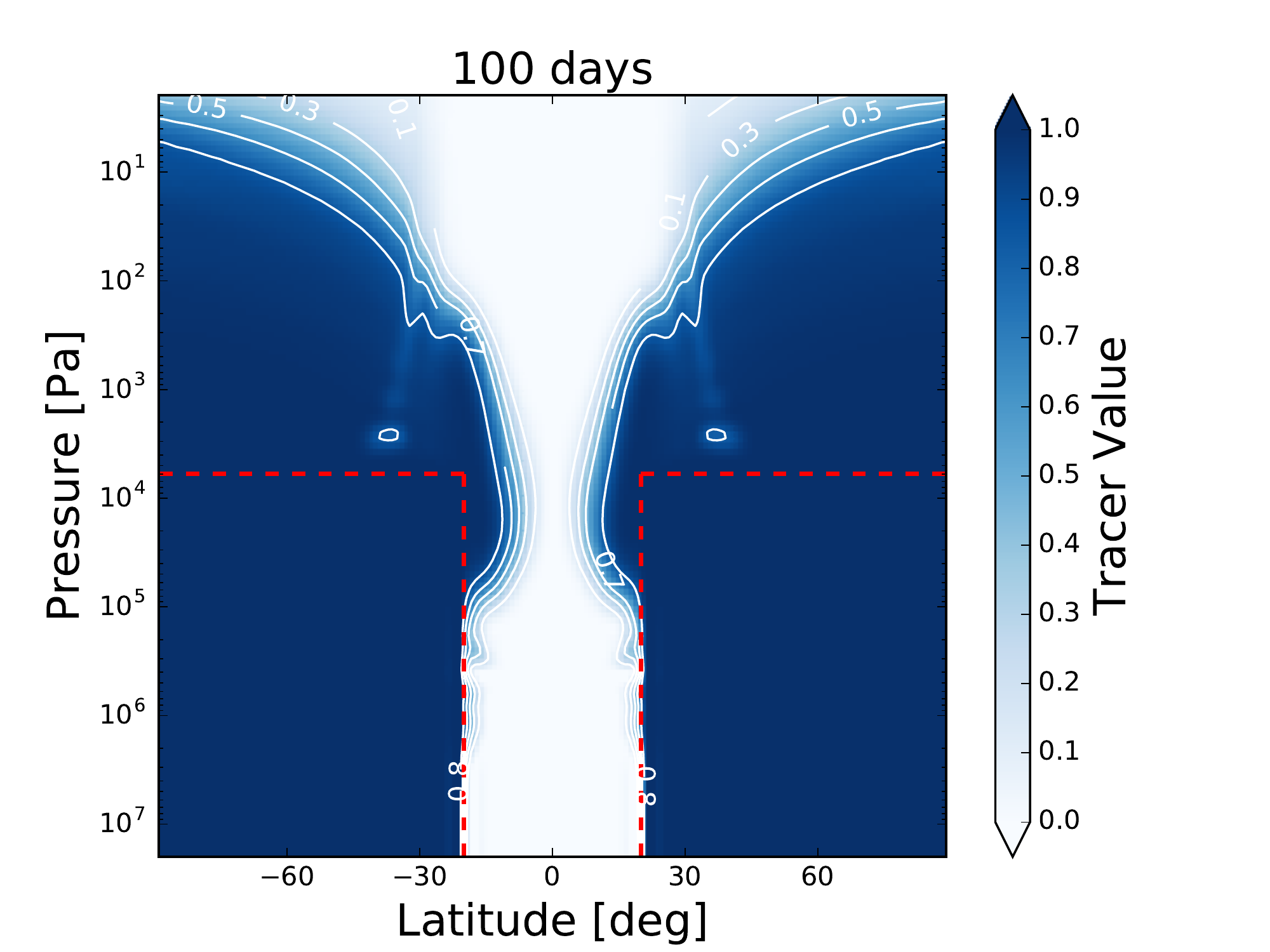}
    \includegraphics[width=0.45\textwidth]{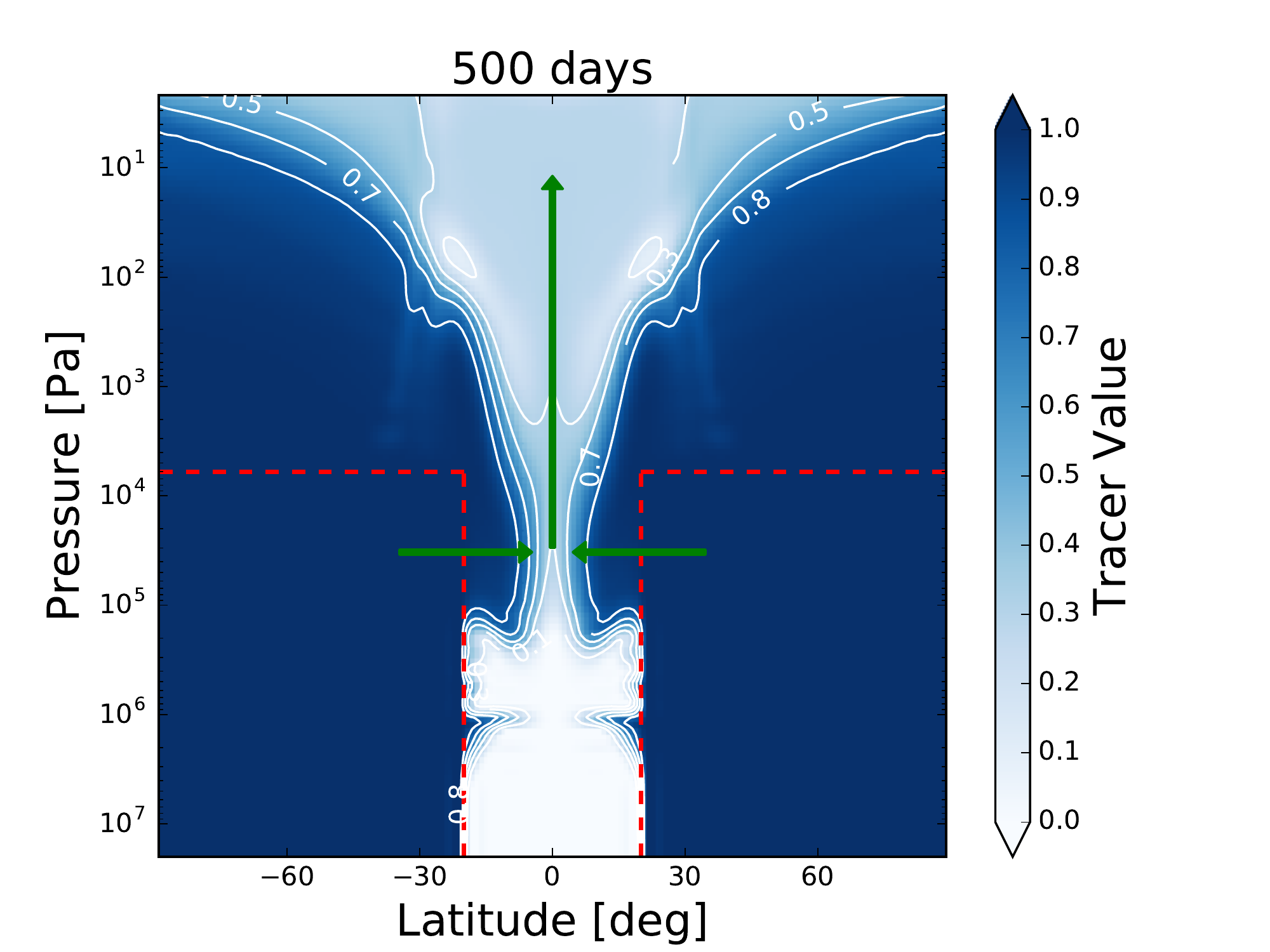}
  \end{center}

  \caption{Snapshots showing the zonal-mean of the tracer value. The regions enclosed by the red dashed lines indicate (approximately) where the tracer is fixed to a value of unity: the `source' regions. Elsewhere, the value of the tracer exponentially decays with a half--life $t_{1/2}=1000$ days. Green arrows illustrate the direction of flow.}
  \label{figure:tracer}
\end{figure*}

To illustrate the transport of material to the equatorial region we perform simulations, initialised from the equilibrium simulation at 1000\,days, that include a tracer which is fixed at unity in a `source' region ($20^{\circ} > \left|\phi\right| > 90^{\circ}$ and $P\gtrsim10^4$) and initialised to zero elsewhere. As the tracer departs the `source' region, its value decays exponentially to zero with time (with a half--life of 1000\,days). \cref{figure:tracer} shows snapshots from this simulation, revealing transport from $\left|\phi\right| > 20^{\circ}$ into the equatorial region, over a timescale of several hundred days, due to equatorward meridional winds (\cref{figure:winds_temp}), and subsequent upwelling. This experiment highlights the importance of the combined effect of horizontal and vertical transport that can only be captured using 3D models.

% TRACER

%%%%%%%%%%%%%%%%%%%%%%%%
% SIMULATED OBSERVATIONS
%%%%%%%%%%%%%%%%%%%%%%%%
\section{Simulated observables}

\begin{figure}
  \center
   \includegraphics[width=0.5\textwidth]{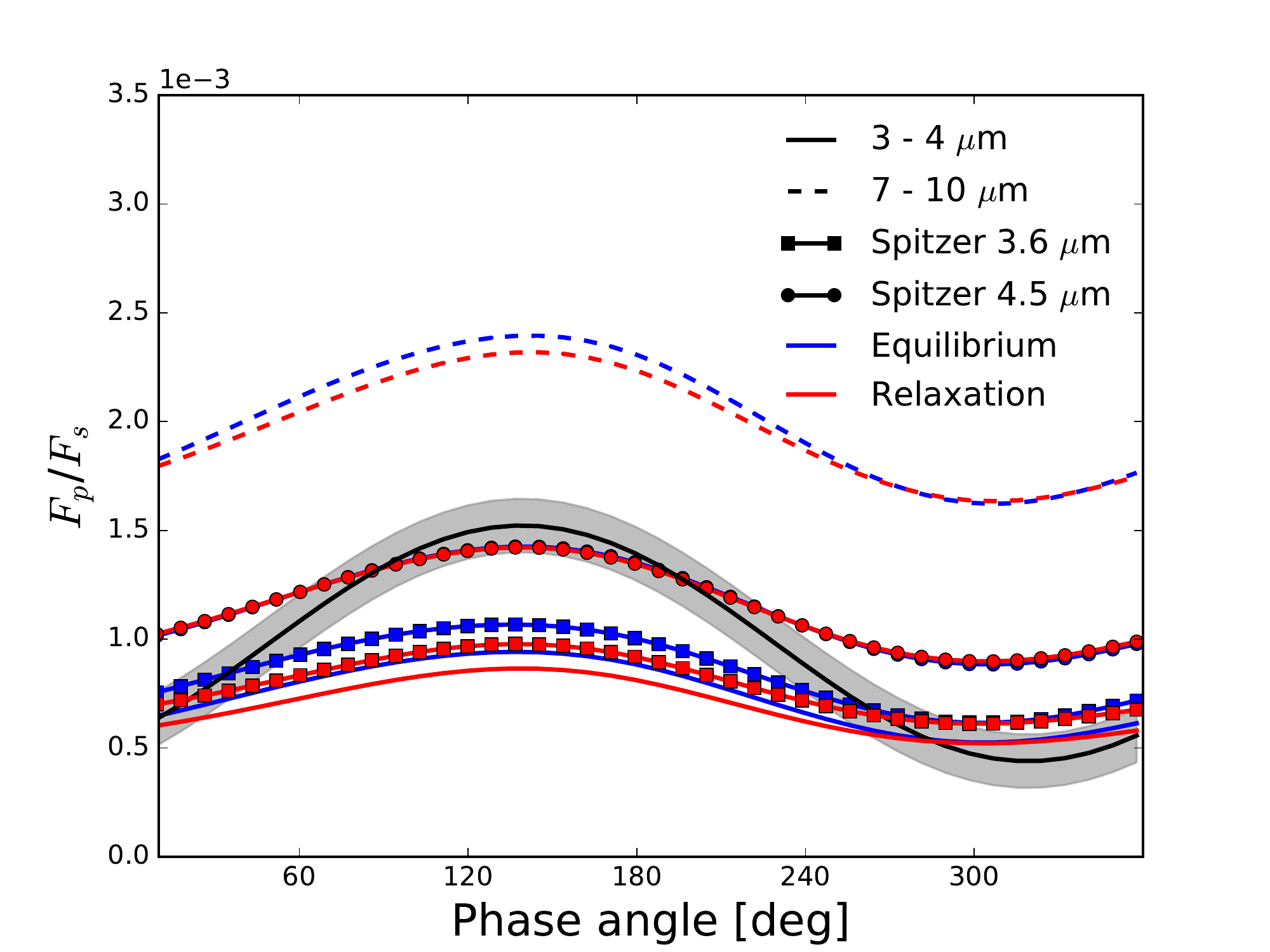} \\
    \includegraphics[width=0.5\textwidth]{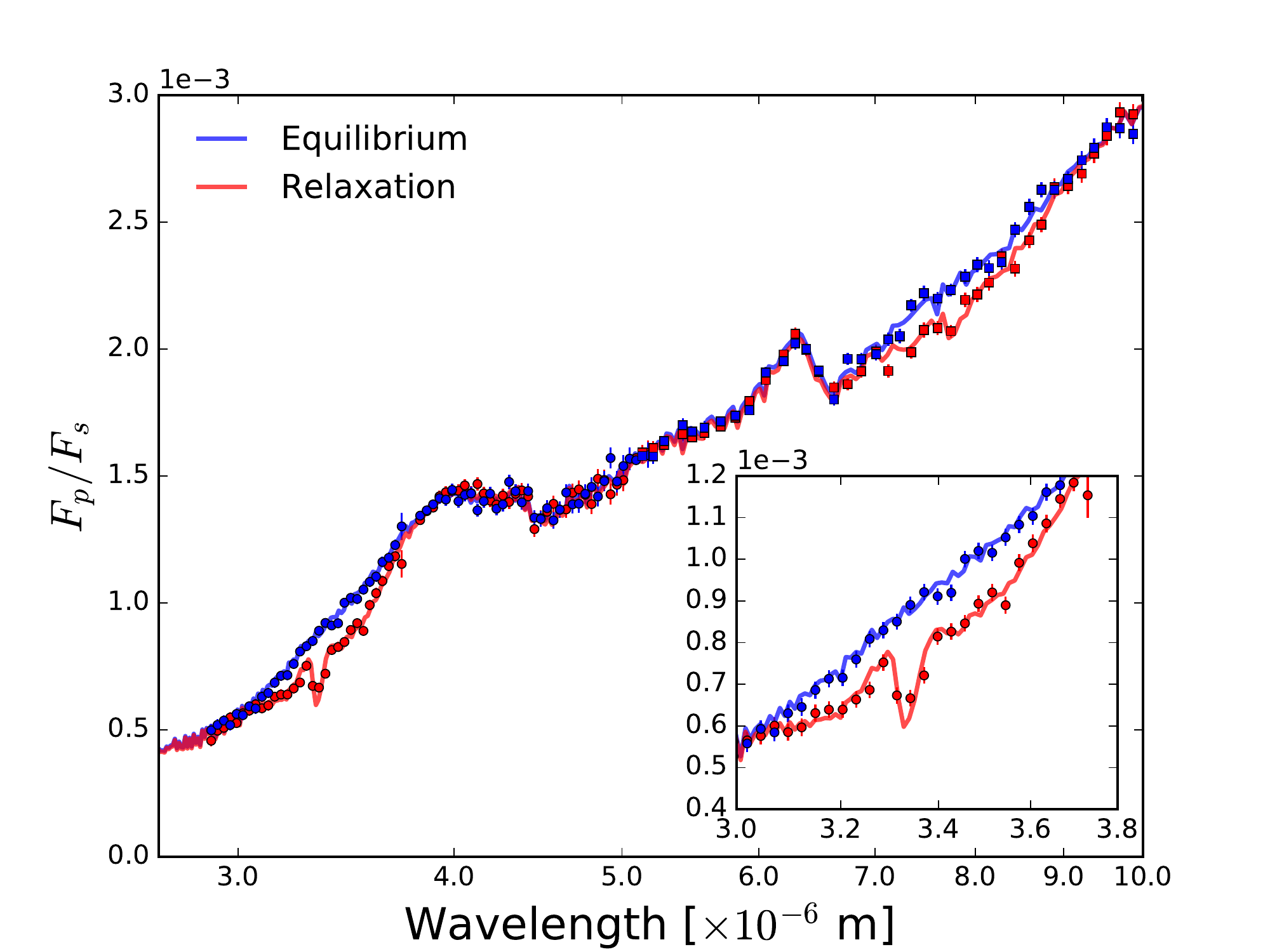} \\
    \includegraphics[width=0.5\textwidth]{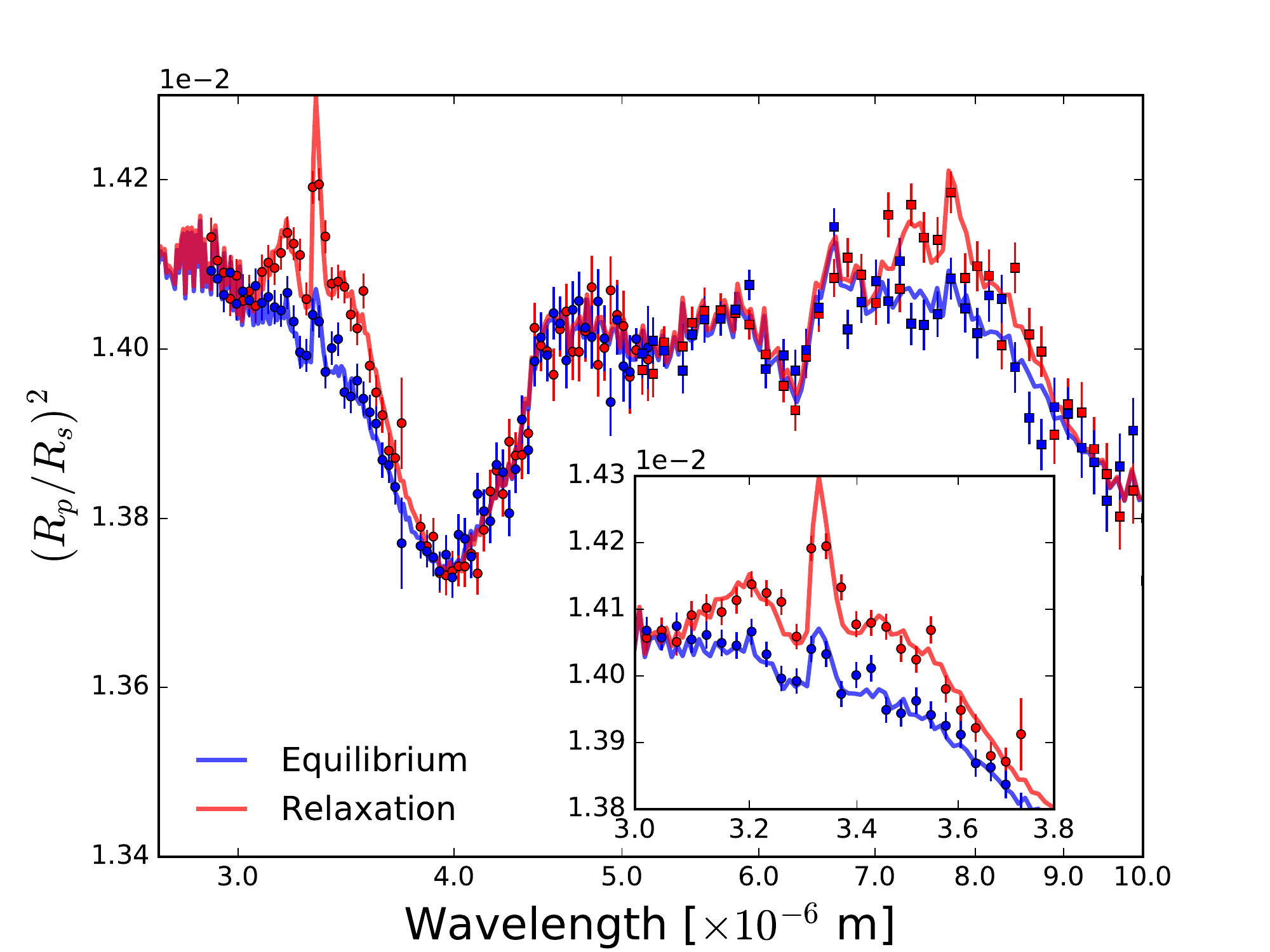} 
\caption{{\it Top}: emission phase curves in several spectral bands. The observed 4.5\,{\textmu m} {\it Spitzer}/IRAC channel curve \citep{ZelLK14} is included (black) with 1$\sigma$ uncertainty. {\it Middle and Bottom}: secondary eclipse emission and transmission spectra with PandExo simulated observations for the NIRSpec G395H (circles) and MIRI LRS (squares) modes, binned to a resolution of $R\sim60$ and $R\sim30$, respectively.}
\label{figure:obs}
\end{figure}

\cref{figure:obs} shows the synthetic observations from our simulations, derived directly from the 3D model: the emission phase curve in several spectral regions, and secondary eclipse and transmission spectra. The observed 4.5\,{\textmu m} phase curve \citep{ZelLK14}, and simulated JWST observations \citep[using PandExo][]{BatMP17}, are both overlaid. The PandExo simulations were performed for the NIRSpec G395H and MIRI LRS modes, using a single eclipse with equal in to out of transit observation time, noise floor of 50\,ppm, detector saturation of 80\,\% full well and stellar and planetary parameters from the TEPCAT database\footnote{\url{http://www.astro.keele.ac.uk/jkt/tepcat/}}. All instrument related parameters were kept at the PandExo defaults.

We find a negligible difference between the emission of the equilibrium and relaxation simulations within the 4.5\,{\textmu m} {\it Spitzer}/IRAC channel, suggesting that gas-phase non-equilibrium chemistry is unlikely to explain the model--observation discrepancy.

More prominent signatures of wind-driven chemistry are apparent within the 3--4\,{\textmu m} and 7--10\,{\textmu m} spectral regions (including the 3.6\,{\textmu m} {\it Spitzer}/IRAC channel), corresponding to methane absorption features. Clear differences between the equilibrium and relaxation simulations can be seen in both transmission and emission, particularly between 3--4\,{\textmu m} for the NIRSpec G395H mode on the James Webb Space Telescope. The decreased flux and increased transit depth both result from enhanced methane abundances that increase the opacity in these spectral regions.

%%%%%%%%%%%%%%%%%%%%%%%%
% CONCLUSIONS
%%%%%%%%%%%%%%%%%%%%%%%%
\section{Discussion and Conclusions}

We have developed the first fully-consistent 3D dynamics, radiative transfer and chemistry model applied to exoplanets. We have focussed on the effect of wind-driven advection on methane, carbon monoxide and water, the major gas-phase absorbers.

Using a chemical relaxation scheme, based on \citet{CooS06}, we have shown that methane is generally enhanced above what is expected from chemical equilibrium, for the specific case of HD~209458b. This is directly opposite to the trend found by previous studies \citep{CooS06,AguPV14}. We find that methane is homogenised, horizontally and vertically, for pressures less than $10^4$ Pa, through a combination of horizontal and vertical transport.

\citet{CooS06} conclude that quenching in the vertical direction is more important than in the horizontal. They argue that vertical quenching transports gas from high pressures, where carbon monoxide is favoured over methane, ultimately leading to a horizontally uniform composition for lower pressures. In contrast, our simulations show that {\it a combination of horizontal and vertical quenching} leads to an increase in the methane abundance, compared with chemical equilibrium.

When using a similar temperature relaxation scheme, rather than deriving the heating rates via radiative transfer calculations, our model gives very similar results to \citet{CooS06}. The temperature relaxation scheme used by \citet{CooS06} results in a significantly warmer dayside and cooler nightside compared with the radiative transfer method \citep{Showman2009,AmuMB16}; the larger day-night contrast then drives a faster equatorial jet. Different locations of the quench points, and different equilibrium abundances at the quench points, due to differing temperature and wind velocity fields, explain the contrasts between our results and those of \citet{CooS06}.

We also find significant differences with \citet{AguPV14} who concluded that a combination of vertical and zonal transport leads generally to a decrease in the methane abundance, compared with chemical equilibrium. While our results suggest, alternatively, that 3D transport acts to increase the methane abundance above equilibrium, the approach of our model and theirs is significantly different. 

\citet{AguPV14} take a highly simplified approach to horizontal transport (solid-body rotation around the equator) but include a full chemical kinetics scheme. On the other hand, our present model considers 3D transport due to the resolved wind but with a highly simplified scheme. We do not expect the uncertainties inherent to the chemical relaxation method to be the reason for the opposite trend compared with \citet{AguPV14}. We suspect that their approximate treatment of horizontal and vertical advection are the main reasons for our opposing results. To confirm this, and to better understand the opposite trends predicted by the two methods, requires coupling a chemical kinetics scheme to a 3D dynamical model, a work which is under progress.

Our results clearly demonstrate the need to include 3D dynamical processes when considering the effect of transport on the composition of exoplanet atmospheres. While vertical quenching dominates over a large pressure range, horizontal transport is important for higher pressures and sets the vertically quenched mole fraction of methane, in the equatorial region. Such processes simply cannot be included consistently in 1D, or even 2D, atmosphere models.

We found small ($\sim\,1\%$) changes to the wind velocity and temperature structure due to the interaction between the composition, the radiative transfer and, subsequently, the dynamics. Interaction between the wind-driven chemistry and the thermal and dynamical structure may be more important for cooler atmospheres, which are closer to the CO=CH$_4$ equilibrium profile, or planets with above solar metallicity, where important absorbers such as carbon dioxide and hydrogen cyanide are more abundant.

We find that wind-driven advection of methane, carbon monoxide and water is unlikely to explain the model--observation discrepancy in the 4.5\,{\textmu m} {\it Spitzer}/IRAC channel emission phase curve. Alternative explanations may involve clouds or hazes \citep{LinMB18} or non-solar abundances \citep{KatSF14,DruMB18} which can alter the dynamics, thermal structure and opacity. 

We find signatures of wind-driven chemistry in the spectral regions 3--4\,{\textmu m} and 7--10\,{\textmu m} due to enhanced methane absorption. These wavelength regions will be accessible with the upcoming James Webb Space Telescope.

Our simulations includes advection due to the resolved, large-scale wind. However, we do not account for unresolved motions which may result in additional transport. An important future work would be to include and quantify this process. We also neglect photochemical processes. However, the main effects of photochemistry are usually restricted to pressures less than 1 Pa \citep[e.g.][]{Moses2011,DruTB16}, beyond the domain of our simulations.

We use a chemical relaxation scheme \citep{CooS06} that parameterises the kinetic interconversion of methane and carbon monoxide, that relies on an approximated timescale. The accuracy of the method therefore depends on the estimation of this timescale. To improve on this we are coupling a chemical kinetics scheme to the UM, which does not rely on such approximations. Coupling a flexible chemical kinetics scheme to a GCM is crucial to enable the study of a larger number of chemical species and a wider range of chemical scenarios, particularly where the chemical relaxation method may not be accurate or relevant. 

In summary, our results indicate that wind-driven chemistry, due to 3D dynamics, leads to a significant increase in the abundance of methane, compared with chemical equilibrium, for HD~209458b--like atmospheres. The increased methane absorption is predicted to significantly effect the emission and transmission spectra, detectable with the James Webb Space Telescope.

\acknowledgements

We are very thankful to the anonymous referee who helped to improve the quality of this letter. BD and DKS acknowledge funding from the European Research Council
(ERC) under the European Unions Seventh Framework Programme
(FP7/2007-2013) / ERC grant agreement no. 336792. NJM is part funded
by a Leverhulme Trust Research Project Grant.  JM and IAB acknowledge
the support of a Met Office Academic Partnership secondment. ALC is funded by an STFC studentship. DSA
acknowledges support from the NASA Astrobiology Program through the
Nexus for Exoplanet System Science. This work used the DiRAC
Complexity system, operated by the University of Leicester IT
Services, which forms part of the STFC DiRAC HPC Facility. This
equipment is funded by BIS National E-Infrastructure capital grant
ST/K000373/1 and STFC DiRAC Operations grant ST/K0003259/1. DiRAC is
part of the National E-Infrastructure.

\bibliographystyle{apj} % style 

\end{document}